\documentclass{article}

\usepackage{arxiv}

\usepackage[utf8]{inputenc} 
\usepackage[T1]{fontenc}    
\usepackage{hyperref}       
\usepackage{url}            
\usepackage{booktabs}       
\usepackage{amsfonts}       
\usepackage{amsmath, amstext, amssymb,bm}
\usepackage{nicefrac}       
\usepackage{microtype}      
\usepackage{lipsum}		
\usepackage{graphicx,subfigure}
\usepackage{natbib}
\usepackage{doi}
\usepackage{psfrag}
\usepackage{array}
\usepackage{float}

\title{Learning Stable Galerkin Models of Turbulence with Differentiable Programming}

\author{ \href{https://orcid.org/0000-0000-0000-0000}{\includegraphics[scale=0.06]{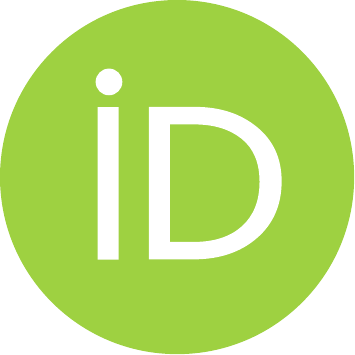}\hspace{1mm}Arvind T.~Mohan}\\
	Computational Physics and Methods Group\\
	Los Alamos National Laboratory\\
	Los Alamos, NM 87545, United States \\
	\texttt{arvindm@lanl.gov} \\
	\And
	\href{https://orcid.org/0000-0000-0000-0000}{\includegraphics[scale=0.06]{orcid.pdf}\hspace{1mm}Kaushik~Nagarajan} \\
	National Aerospace Laboratories\\
	Bengaluru, Karnataka 560 017, India \\
	\AND
	\href{https://orcid.org/0000-0000-0000-0000}{\includegraphics[scale=0.06]{orcid.pdf}\hspace{1mm}Daniel~Livescu} \\
	Computational Physics and Methods Group\\
	Los Alamos National Laboratory \\
	Los Alamos, NM 87545, United States \\
}

\renewcommand{\shorttitle}{\textit{arXiv} Template}

\hypersetup{
pdftitle={Learning Stable Galerkin Models of Turbulence with Differentiable Programming},
pdfsubject={physics.flu-dyn, nlin.CD},
pdfauthor={Arvind T.~Mohan, Kaushik~Nagarajan, Daniel~Livescu},
pdfkeywords={Reduced Order Modeling, Differentiable Programming, Neural Networks},
}

\begin{document}
\maketitle

\begin{abstract}
Turbulent flow control has numerous applications and building reduced-order models (ROMs) of the flow and the associated feedback control laws is extremely challenging. Despite the complexity of building data-driven ROMs for turbulence, the superior representational capacity of deep neural networks has demonstrated considerable success in learning ROMs. Nevertheless, these strategies are typically devoid of physical foundations and often lack interpretability. Conversely, the Proper Orthogonal Decomposition (POD) based Galerkin projection (GP) approach for ROM has been popular in many problems owing to its theoretically consistent and explainable physical foundations. However, a key limitation is that the ordinary differential equations (ODEs) arising from GP ROMs are highly susceptible to instabilities due to truncation of POD modes and lead to deterioration in temporal predictions. In this work, we propose a \textit{differentiable programming} approach that blends the strengths of both these strategies, by embedding neural networks explicitly into the GP ODE structure, termed Neural Galerkin projection. We demonstrate this approach on the isentropic Navier-Stokes equations for compressible flow over a cavity at a moderate Mach number. When provided the structure of the projected equations, we show that the Neural Galerkin approach implicitly learns stable ODE coefficients from POD coefficients and demonstrates significantly longer and accurate time horizon predictions, when compared to the classical POD-GP assisted by calibration. We observe that the key benefits of this differentiable programming-based approach include increased flexibility in physics-based learning, very low computational costs, and a significant increase in interpretability, when compared to purely data-driven neural networks.
\end{abstract}


\section{Introduction}
Nonlinear dynamical systems pervade several natural phenomena, with the most popular example being turbulence in fluids. Turbulent flow governs the efficiency and performance of several engineering systems where drag reduction, noise suppression and improved fuel efficiency are critical parameters of interest. Naturally, controlling turbulence with flow control devices has been an active area of research in recent years, to design better aircraft, vehicles, and energy systems~\citep{gad2001flow,bons2005designing}. However, controlling a chaotic and nonlinear system like turbulence requires making predictions several time instants into the horizon in a fast, cheap and accurate manner. A model satisfying these characteristics is called a reduced order model or ROM ~\citep{noack2011reduced,luchtenburg2010turbulence}, since CFD techniques require computational resources which are prohibitive for on-board control hardware in these applications. ROMs are a crucial component of the overall flow control strategy~\citep{ito1998reduced,ma2011reduced,ravindran2000reduced}, since such predictions are inputs to the feedback control algorithms that operate the actuator by the desired amount, over a designated time interval. Therefore, research into ROMs of turbulence has relied on data-driven strategies and approximations, since speed and efficiency of ROMs take precedence over high-fidelity solutions. One of the most popular strategies in literature involves the Galerkin projection of the Navier-Stokes equations onto truncated orthogonal basis vectors (hereafter termed GP-ROM), such that the projected equations can be solved quickly, while capturing the dominant physics of the flow~\citep{rempfer2000low,rowley2004model,lorenzi2016pod}. The GP-ROM is based on the Navier-Stokes equations and is fast, efficient, with low computational costs and thus can be accommodated with feedback control algorithms. However, current GP-ROMs suffer from several issues of accuracy and stability over long-time horizons, which are required for effective control~\citep{iollo2000stability,balajewicz2012stabilization,amsallem2012stabilization,akhtar2009stability}.

Recent efforts in literature have explored alternatives to GP-ROM with deep learning approaches such as neural networks (NNs). Several variants of NNs exist, with the LSTM architecture~\citep{hochreiter1997long} being popular in turbulence~\citep{mohan2018deep,ahmed2020long,deng2019time,maulik2020non,chattopadhyay2020data} owing to its strengths in time-series prediction. Although NNs have been reported to have considerable promise, they face important limitations to practical use due to their black-box nature and lack of interpretability. In this work, we present a hybrid approach that combines Galerkin Projections with deep learning using \textit{Differentiable Programming} (henceforth referred to as \textit{DiffProg}). DiffProg is a programming paradigm that enables writing software that can be fully differentiated via automatic differentiation~\citep{griewank1989automatic,bartholomew2000automatic,baydin2018automatic}, and optimized using gradient-descent based methods. While the idea behind DiffProg is not new~\citep{bischof1996adifor,yu2013dnad,utke2008openad} it has seen renewed interest due to recent efforts in interpreting deep neural networks (NNs) through the lens of DiffProg~\citep{abadi2019simple,chizat2018lazy,li2018differentiable,wang2018language}, with the most prominent examples being Neural ODEs~\citep{chen2018neural}. The initial successes of NNs in addressing scientific problems has made DiffProg an attractive avenue of research~\citep{innes2019differentiable,hu2019difftaichi,rackauckas2020universal}, since NNs can be viewed as an application of DiffProg to specific learning algorithms. 
In this work, we present an approach called \textit{Neural Galerkin Projection (NeuralGP)} which exploits DiffProg to achieve superior accuracy and long term stability. This is achieved by casting the traditional GP-ROM as a DiffProg problem where neural networks are directly embedded inside the equations, and trained on CFD data. The resulting NeuralGP framework
Section~\ref{sec:GPintro} outlines the Galerkin Projection approach as applied to flow over a cavity. Section~\ref{sec:GPNODE}  provides an introduction to Differentiable programming in the context of ordinary differential equations, and our extensions to incorporate it into Galerkin Projections. The results from this new approach are presented in Section~\ref{sec:results} and compared with the standard GP-ROM approach, where we demonstrate superior accuracy in the predicted POD coefficients and their stability over longer time horizons. Finally, in Section~\ref{sec:conclusion}, we discuss the strengths, novelty and future of adopting DiffProg in ROMs of turbulence.

\section{Galerkin Projections for ROMs}
\label{sec:GPintro}
The first step in GP is to pre-process the flow data with Proper Orthogonal Decomposition (POD).
The basic idea of a POD is to decompose the flow field into energy ranked coherent structures represented by mathematical modes~\citep{berkooz1993proper,smith2005low}. Let $H$ denote any Hilbert space with an inner product denoted by $\left(.,.\right)$ and the norm induced by the inner product denoted by $\left\Vert.\right \Vert$.  In the fluid dynamic context, the space $H$ consists of functions defined on some spatial domain $\Omega \subset R^d$ ($d$ is the spatial dimension) on which the fluid evolves, usually $H = L^{2}(\Omega)$.  Let $\bm{q}(\bm{x},t)$ be any vector of flow variables,  where $\bm{x}$ and $t\in[0,T]$ denote respectively the spatial variables and the time of the numerical simulation or experimental measurements. We also assume a discrete sampling of our data in time and space.  We seek an expansion for $\bm{q}$ of the form
\begin{equation} 
 \bm{q}(\bm{x},t) =  \bm{\overline{q}}(\bm{x})+\sum_{i=1}^{n} a_i(t)\bm{\phi_i}(\bm{x}),
 \label{PODeqn}
\end{equation}
where $\bm{\overline{q}}$ denotes the average of the $n$ flow snapshots, such that the flow is decomposed into an averaged component and a fluctuating component. The fluctuating component is further represented by a POD expansion, which contains $n$ spatial modes $\bm{\phi_i}$ and their temporal coefficients $a_i(t)$. We truncate the number of POD modes and coefficients in Eqn.~\ref{PODeqn} by only retaining modes with the highest norm. This truncation serves as a lossy dimensionality reduction tool, since we represent the flow field by just a few modes. These modes are assumed to represent the energetically dominant flow structures in the data, while the truncated modes account for energetically less important flow features, and consequently can be neglected. The reader is directed to the Appendix for a brief theoretical background of POD. 

The ROM is obtained by performing a Galerkin projection of the governing dynamics (in this case the Navier-Stokes equations) onto the POD spatial modes $\phi_i$.  Two different ROMs for cavities have been proposed in \cite{gloerfelt2008compressible}, one using the full compressible equations and the other with an isentropic approximation.  In this work, we use the isentropic equations for compressible flows \citep{rowley2004model}, which is a valid physical representation for flows at moderate Mach number. Scaling the stream-wise and span-wise component of velocities $u,\,v$ by the free stream velocity $U_\infty$, the local sound speed $c$ by the ambient sound speed $c_\infty$, the lengths by the cavity depth $D$, and time by $D/U_\infty$, the equations are given by
\begin{eqnarray}
   u_t + uu_x+vu_y+\frac{1}{M^2}\frac{2}{\gamma -1} cc_{x} = \frac{1}{Re}(u_{xx}+u_{yy}) \nonumber \\
   v_t + uv_x+vv_y+\frac{1}{M^2}\frac{2}{\gamma -1} cc_{y} = \frac{1}{Re}(v_{xx}+v_{yy}) \nonumber \\
   c_{t} + uc_{x}+vc_{y}+\frac{\gamma -1}{2}c(u_x+v_y) =0
  \end{eqnarray}
where $M=U_\infty/a_\infty$ is the Mach number and $Re=U_\infty D/\nu$ is the Reynolds number.  If we denote $\bm{q} = (u,v,c)$  the vector of flow variables,  the above equations can be recast as
\begin{equation}\label{newrom}
\bm{\dot{q}} = \frac{1}{Re}\bm{L}(\bm{q})+\frac{1}{M^2}\bm{Q_1}(\bm{q},\bm{q})+\bm{Q_2}(\bm{q},\bm{q})        \qquad \qquad \hbox{with}
\end{equation}

\begin{eqnarray*}
\bm{L}(\bm{q})  =
\begin{bmatrix}
	u_{xx}+u_{yy} \\
	v_{xx}+v_{yy}\\
	0
\end{bmatrix}
, &\hspace{0.0cm}
\bm{Q_1}(\bm{q^1},\bm{q^2})  = -\frac{2}{\gamma-1}
\begin{bmatrix}
	c^1c_{x}^2 \\
        c^1c_{y}^2 \\
         0
\end{bmatrix}
, &\hspace{0.0cm}
\bm{Q_2}(\bm{q^1},\bm{q^2})  = -
\begin{bmatrix}
	u^1u_{x}^2+v^1u_{y}^2\\
	u^1v_{x}^2+v^1v_{y}^2\\
	u^1c_{x}^2+ v^1c_{y}^2 \\
      +\frac{\gamma-1}{2}c^1(u_{x}^2+v_{y}^2)
\end{bmatrix}
\end{eqnarray*}

To obtain the ROM by means of a GP we define an inner product on the state space as
$$\left(\bm{q_1},\bm{q_2} \right)_\Omega = \int_\Omega (u_1u_2+v_1v_2+\frac{2\alpha}{\gamma-1}c_1c_2)\,d\Omega$$
where $\alpha$ is a constant and $\gamma$ is the ratio of specific heats. In this work, we choose the value of $\alpha=1$, which gives
the definition of stagnation enthalpy while calculating the norm. 	The above definition ensures the stability of the origin of the attractor as demonstrated by \cite{rowley2004model}.  We use expansion in Eqn.~\ref{PODeqn} to perform the GP of the isentropic equations onto the first $n\ll N_{POD}$ spatial eigenfunctions, to obtain the ROM as     

\begin{eqnarray}  
\nonumber \dot a_i^R(t) & = & \frac{1}{Re}C_i^1+\frac{1}{M^2}C_i^2+C_i^3+
\nonumber \displaystyle\sum_{j=1}^n\left(\frac{1}{Re}L_{ij}^1+\frac{1}{M^2}L_{ij}^2+L_{ij}^3\right)a_j^R(t) \\
 &+&
\nonumber \displaystyle\sum_{j,k=1}^n\left(\frac{1}{M^2}Q_{ijk}^1+Q_{ijk}^2\right)a_j^R(t)a_k^R(t)\\
 \nonumber & =& C_i + \displaystyle\sum_{j=1}^n L_{ij} a_j^R(t) + \displaystyle\sum_{j,k=1}^n Q_{ijk} a_j^R(t)a_k^R(t) \\  \nonumber &=&  f_i(\underbrace{C_i,\bm{L_i},\bm{Q_i}}_{\bm{y}},\bm{a}^R(t)) \\
           &=&  \bm{f}(\bm{y},\bm{a}^R(t) )
\label{rom}
\end{eqnarray}
where $f_i$  is a  polynomial of degree 2 in $\bm{a}^R$ and the coefficients of the projection are given by
\begin{eqnarray*}
\begin{matrix}
	 C_i^1 = \left( \bm{\phi_{i}},\bm{L}(\bm{\overline{q}})\right)_\Omega \\
         C_i^2 = \left( \bm{\phi_i}, \bm{Q_1}(\bm{\overline{q}},\bm{\overline{q}})\right)_\Omega \\
         C_i^3 = \left( \bm{\phi_i}, \bm{Q_2}(\bm{\overline{q}},\bm{\overline{q}})\right)_\Omega
\end{matrix}
\hspace{0.1cm}
\begin{matrix}
        L_{ij}^1 = \left( \bm{\phi_{i}}, \bm{L}(\bm{\phi_{j}}) \right)_\Omega \\
	L_{ij}^2 = \left( \bm{\phi_{i}},\bm{Q_1}(\bm{\overline{q}},\bm{\phi_{j}})+\bm{Q_1}(\bm{\phi_{j}},\bm{\overline{q}}) \right)_\Omega \\
        L_{ij}^3 = \left( \bm{\phi_{i}}, \bm{Q_2}(\bm{\overline{q}},\bm{\phi_{j}})+\bm{Q_2}(\bm{\phi_{j}},\bm{\overline{q}})\right)_\Omega
\end{matrix}
\hspace{0.1cm}
\begin{matrix}
	Q_{ijk}^1 = \left( \bm{\phi_{i}},\bm{Q_1}(\bm{\phi_{j}},\bm{\phi_{k}}) \right)_\Omega \\
        Q_{ijk}^2 = \left( \bm{\phi_{i}},\bm{Q_2}(\bm{\phi_{j}},\bm{\phi_{k}}) \right)_\Omega
\end{matrix}
\end{eqnarray*}

\section{Differentiable Programming for Galerkin ROM}
\label{sec:GPNODE}
\subsection{Overview}
\label{sec:methods:NODE}
In recent years, differentiable programming has taken a center-stage in scientific ML applications, such as interpreting neural networks from the perspective of differential equations and physics, especially as applied to dynamical systems~\citep{cessac2010view}. A popular example of this is the \textit{Neural Ordinary Differential Equations} (NODE)~\citep{chen2018neural}. To briefly understand the key idea underpinning NODE (and consequently DiffProg), consider a derivative $g$ that is defined as a function of some independent variables. Let us assume that the true form of $g$, denoted by $\hat{g}$, is unknown, since this scenario is commonly observed in several scientific and real-world problems. When  formulating this as a deep learning problem, $g$ is the function that is represented by a NN, which upon training and convergence should approximate the true function $\hat{g}$. In essence, we are approximating the derivative of an unknown function with a NN. 
\begin{equation}
  \frac{dy}{dx} = g(x,y)  
  \label{derivative}
\end{equation}
In order to appreciate the connection between this derivative function and NODE, we briefly review the standard \textit{residual network} and \textit{recurrent network} architectures, which have been popular in deep learning. These NNs are composed by stacking upon several neural
layers known as \textit{hidden layers}. The outputs of these layers are called hidden states $h$, and can be written in the form
\begin{equation}
    \mathbf{h}_{t+1} = \mathbf{h}_{t} + f(\mathbf{h}_{t},\theta_{t})
\end{equation}
where time $t \in \{ 0...T\}$ and $\mathbf{h}_{t} \in \mathcal{R}^{D}$, while $\theta_t$ are the trainable parameters in the NNs. The fundamental premise of NODE is that the iterative updates to $\mathbf{h}$ can be thought of as an Euler discretization of a continuous transformation~\citep{haber2017stable,chang2018reversible,behrmann2019invertible}. This idea is further exploited when we add more layers and take smaller steps, such that in the limit, the expression above can be written in continuous dynamics as an ODE:
\begin{equation}
    \frac{d \mathbf{h}(t)}{dt} = f(\mathbf{h}(t), t,\theta).
\end{equation}
This ODE now represents the rate of change of the hidden states (and hence, their dynamics) as a parameterized function $f$ with parameters $t$ and $\theta$. Since time $t \in \{ 0...T\}$ in a residual or recurrent NN is ``unrolled" through several hidden layers, solving for $\mathbf{h}(t)$ now requires solving an ODE with initial condition $\mathbf{h}(0)$, which is input at $t\,=\,0$. Consequently, the solution of the ODE at time T is the solution of output at the final layer $\mathbf{h}(T)$. Framing the NN problem as an ODE has a significant advantage since solving ODE problems is a mature area of scientific computing, with several ODE solvers available that can solve these equations in an efficient, accurate manner. Several adaptations in modern ODE solvers exist, such as keeping the solution error bounded across time and explicit control over the level of accuracy by changing the numerical method. Importantly, the computational cost of these ODE solvers is low enough that they can be quickly solved on commodity computing hardware. This is in contrast to residual and recurrent NNs, which require progressively higher computational and memory requirements as $T$ increases, since this leads to a corresponding increase in $\mathbf{h}(t)$. Recurrent NNs are especially notorious for their significant memory requirements when unrolling the hidden states for long periods of time, and need dedicated high performance computing resources. A comparison of accuracy and speed between NODE and recurrent NNs can be found in~\cite{maulik2020time}.

Though the benefits are significant, a key roadblock exists in practice. Since $f$ is now approximated by a NN, analogous to Eqn.~\ref{derivative}, incorporating a NN inside an ODE solver requires training it with a backpropagation strategy; while simultaneously solving the ODE with numerical strategies like the Euler method. It is apparent that this requires \textit{backpropagation through the ODE solver}, which amounts to gradient computation via reverse-mode automatic differentiation (AD). This is an important bottleneck since employing the standard reverse-mode approach commonly used in NNs to an ODE, results in high memory costs and is slow, in addition to lack of guaranteed accuracy limits. The key contribution by~\cite{chen2018neural} circumvents this by computing the gradients via the adjoint sensitivity method of~\cite{boltyanskiy1961theory}. To illustrate its operation, consider a scalar-valued loss function $L$, which is typical in computing reverse-mode AD with chain rule. The loss is the output of the predictions by the ODE solver such that
\begin{equation}
    L(\mathbf{z}(t_{0})) \,=\, L \left( \mathbf{z}(t_{0}) + \int_{t_{0}}^{t_{1}} f(\mathbf{z}(t),t,\theta)dt \right) \,=\, L(\mathrm{ODESolve}(\mathbf{z}(t_{0}),f,t_{0},t_{1},\theta))
\end{equation}
The learning problem requires gradients to be computed with respect to parameters $\theta$, to minimize loss L. Since these gradients depend on the hidden state $z(t)$, the adjoint to be computed can be written as $ \mathbf{a}(t) \,=\, \frac{\partial L}{\partial z(t)}$, which can then be represented as another ODE
\begin{equation}
    \frac{d\mathbf{a}(t)}{dt} \,=\, - \mathbf{a}(t)^{T} \frac{\partial f(\mathbf{z}(t),t,\theta)}{\partial \mathbf{z}} 
\end{equation}
Finally, gradients of L with respect to parameters $\theta$ are computed via
\begin{equation}
    \frac{dL}{d \theta} \,=\, - \int_{t_{1}}^{t_{0}} \mathbf{a}(t)^{T} \frac{\partial f(\mathbf{z}(t),t,\theta)}{\partial \theta} dt 
\end{equation}
We conclude this brief description of NeuralODE with its underlying philosophy of blending adjoint methods in ODEs with backpropagation, for interoperability between neural networks and ODEs. We encourage the reader to the works of~\cite{chen2018neural,rubanova2019latent,zhang2019approximation,yan2019robustness} and~\cite{rackauckas2020universal} for a detailed treatment of the mathematical intricacies and implementations of NODE, as it is beyond the scope of this paper.

\subsection{Our Approach: Neural Galerkin Projections}
\label{sec:methods:NGP}
With a brief description of the Neural ODE approach above, we now re-direct our attention to the GP formulation described in Section~\ref{sec:GPintro}. The equations to compute GP with POD modes can be written in ODE form as
\begin{equation}
\dot a_i^R(t) =  C_i + \displaystyle\sum_{j=1}^{N} L_{ij} a_j^R(t) + \displaystyle\sum_{j,k=1}^{N} Q_{ijk} a_j^R(t)a_k^R(t)
\label{gpEqn}
\end{equation}
The GP formulation gives rise to $C$ (Constant), $L$ (Linear) and $Q$ (Quadratic) terms, that are determined via an optimization and calibration process \textit{from the data}. We remark that the data-driven nature of Eqn.~\ref{gpEqn} while being structured as an ODE, makes it an excellent candidate to be instead cast as a DiffProg problem. With DiffProg, the ``unknown" operators in the equations are represented by a NN of appropriate dimensionality, and the adjoint-based backpropagation learns this NN from the data. Specifically, we have seen that $C$ and $L$ are the coefficients that have significant impact on the accuracy and long-time stability of GP~\citep{nagarajan2013development}, and need considerable manual intervention in the form of calibration. Therefore, our intention is to automate discovery of these coefficients with DiffProg, and directly using the $a_i^R(t)$ from the DNS as the training data. This is done by substituting the coefficients $C$ and $L$ with individual NNs. The NNs take as input a vector $f(q)$, which is dependent on the flow variables $q$ which can include velocity, density and pressure. The resulting NeuralGP setup is now written as
\begin{align}
    C^{\theta} \,&=\, NN_{\theta}^{C}(f(q)^{C}) \\
    L^{\theta} \,&=\, NN_{\theta}^{L}(f(q)^{L}) \\
    \dot a_i^R(t) \, &= \, C_{i}^{\theta} + \displaystyle\sum_{j=1}^{N} L_{ij}^{\theta} a_j^R(t) + \displaystyle\sum_{j,k=1}^{N} Q_{ijk} a_j^R(t)a_k^R(t) 
    \label{neuralGP}
\end{align}
Where $C^{\theta}$ refers to the unknown $C$ that is now parameterized with a NN having parameters $\theta$, and similarly for $L^{\theta}$. Thus, in the forward pass of the NeuralGP, both NNs are provided their respective inputs $f(q)$ and predict the $C$ and $L$ coefficients, upon which Eqn.~\ref{neuralGP} is solved with standard ODE solvers to make predictions. In the backward pass, the adjoint method described above trains the NNs $C^{\theta}$ and $L^{\theta}$. At this juncture, its useful to discuss some salient advantages of the NeuralGP formulation by comparing Eqn.~\ref{gpEqn} to Eqn.~\ref{neuralGP}

\begin{figure}
    \centering
    \includegraphics[width=12cm]{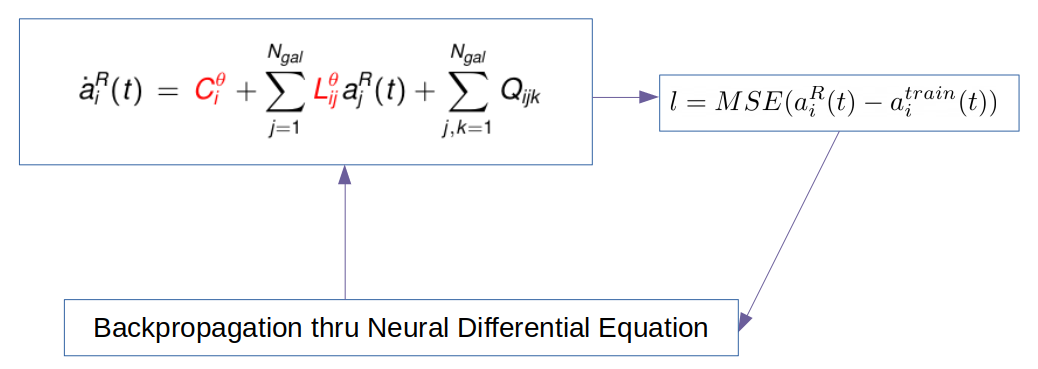}
    \caption{Schematic of NeuralGP training with embedded NNs in the Galerkin Projection ODE}
    \label{fig:NeuralGPgraphic}
\end{figure}

\begin{itemize}
    \item Both $C_{ij}$ and $L_{ij}$ have their own NN models, but they interact in a \textit{physical manner} governed by the projected Navier-Stokes equations.
    \item DiffProg allows for NN models to be backpropagated seamlessly thru the ODE solver via  adjoint sensitivity analysis and a combination of forward/reverse mode Automatic Differentiation (AD). The resulting abstract, powerful framework allows for arbitrary neural network models to be tightly integrated within the context of the well-known GP equations, which are interpretable.
    \item The choice of the input vector $f(q)$ is flexible and user defined, as it can be an initial guess based on dependency relationship between the coefficients and the flow physics, or even be random matrix - in which case, it acts as an ``input seed" for the NN to make a prediction.
\end{itemize}
These inherent theoretical advantages of NeuralGP imply superior capability to the standard GP and we present evidence of this in the results below.

\section{Results}
\label{sec:results}
The objective of this work is to demonstrate the performance, accuracy and effectiveness of the proposed NeuralGP method. The first step in this pursuit is to have benchmark results obtained with the standard GP approach, on the same dataset. In this section, we first construct a GP-ROM using the standard approach found in literature~\citep{rowley2004model,nagarajan2013development}, using a combination of optimization and calibration. 
\begin{figure}
\centering
\psfrag{x}[c][c][0.8]{$\bm{\hat{e}}_x$}
\psfrag{y}[c][l][0.8]{$\bm{\hat{e}}_y$}
\psfrag{o}[c][c][0.8]{$0$}
\psfrag{l}[c][c][0.8]{$L_e=2D$}
\psfrag{d}[c][c][0.8]{$D$}
\psfrag{u}[c][c][0.8]{$U_\infty$}
\psfrag{b}[c][c][0.8]{"Buffer zone"}
\psfrag{br}[c][c][0.8][90]{"Buffer zone"}
\psfrag{d1}[c][c][0.8]{$6D$}
\psfrag{d2}[c][c][0.8]{$2.5D$}
\psfrag{d3}[c][c][0.8]{$10 D$}
\centerline{\includegraphics[width=0.5\linewidth]{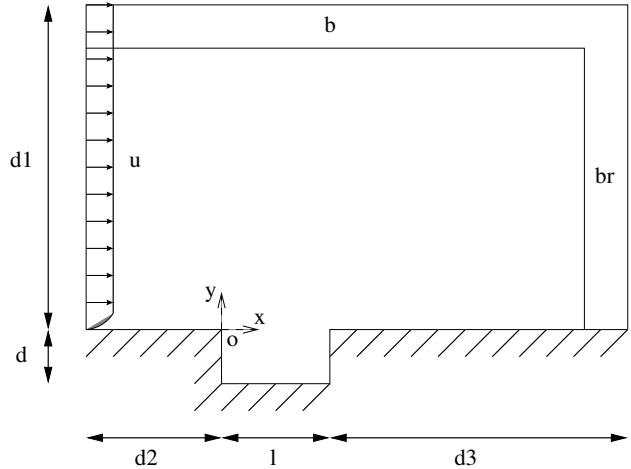}}
\caption{Cavity configuration and computational domain.}
\label{fig:cavity}
\end{figure}
The dataset chosen is a 2D DNS dataset of cavity flow as shown in Fig.~\ref{fig:cavity}. The cavity is of an $L_e/D$ ratio of $2$.  The flow is initialised by a laminar boundary layer so as  to have a  thickness  of $\delta/D= 0.28$ at the leading edge of the cavity.  The Reynolds number of the flow based on the cavity depth is $1500$ and the flow Mach number is $0.6$. The equations of Navier-Stokes are discretised using a $4^{th}$ order accurate scheme in time and space. This flow demonstrates vortex shedding from the edge of the cavity and is a canonical abstraction of engineering applications in aeroacoustics. This vortex shedding has several instabilities which trigger a strong acoustic response, and are known as Rossiter modes. Suppression and manipulation of these modes have been a focus of several research efforts in aerodynamics, with a famous example being the payload bay of an aircraft which has a strong acoustic signature during flight~\citep{stanek2000control,levasseur2008unstructured,williams2007supersonic}. Therefore, this case acts as a good representative of several engineering flows that can benefit from feedback control problems. The differentiable programming computations are performed in the \textit{Julia} programming language~\citep{bezanson2017julia} with the \textit{DiffEqFlux}~\citep{rackauckas2019diffeqflux} package, as it offers tight integration between adjoint solvers in ODEs and automatic differentiation required to train neural networks.

\subsection{Galerkin Projection ROMs of Cavity Flow}
\label{sec:results:GP}
\begin{figure}
\psfrag{x}{ $k$ mode}
\psfrag{y}{ $\log10 \ \textrm{Re}(\lambda_k)$}
\psfrag{z}{ RIC in $\%$}
\psfrag{leg1}{ $\log10 \ \textrm{Re}(\lambda_k)$ }
\psfrag{leg2}{RIC in $\%$}
\centerline{\includegraphics[width=0.5\linewidth,angle=0]{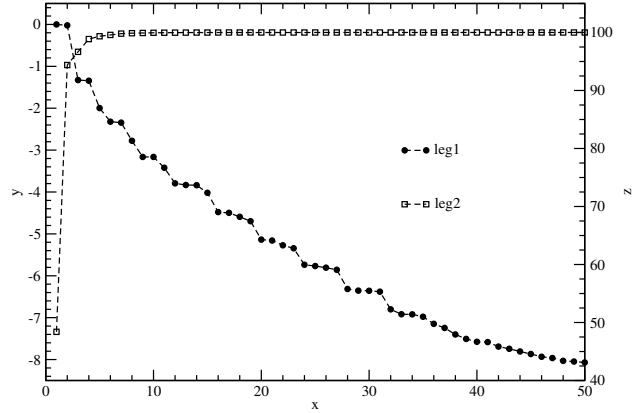}}
\caption{Real part of the eigen values   and  Relative Information Content (RIC)}
\label{figcicp:ric1}
\end{figure}

\begin{figure} 
\psfrag{x}{$t$}
\psfrag{y}{$a_i$}
\psfrag{a1}{$a_1$}
\psfrag{a2}{$a_2$}
\psfrag{a3}{$a_3$}
\psfrag{a4}{$a_4$}
\psfrag{a5}{$a_5$}
\psfrag{a6}{$a_6$}
\centerline{\includegraphics[width=0.5\linewidth,angle=0]{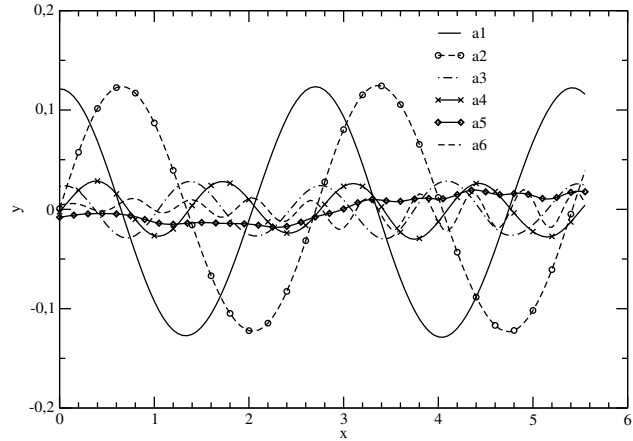}}
\caption{First 6 POD temporal coefficients}
\label{figcicp:temporalcoeff}
\end{figure}

\begin{figure}
\centering
\subfigure[mode 1]
{\includegraphics[width=0.45\textwidth]{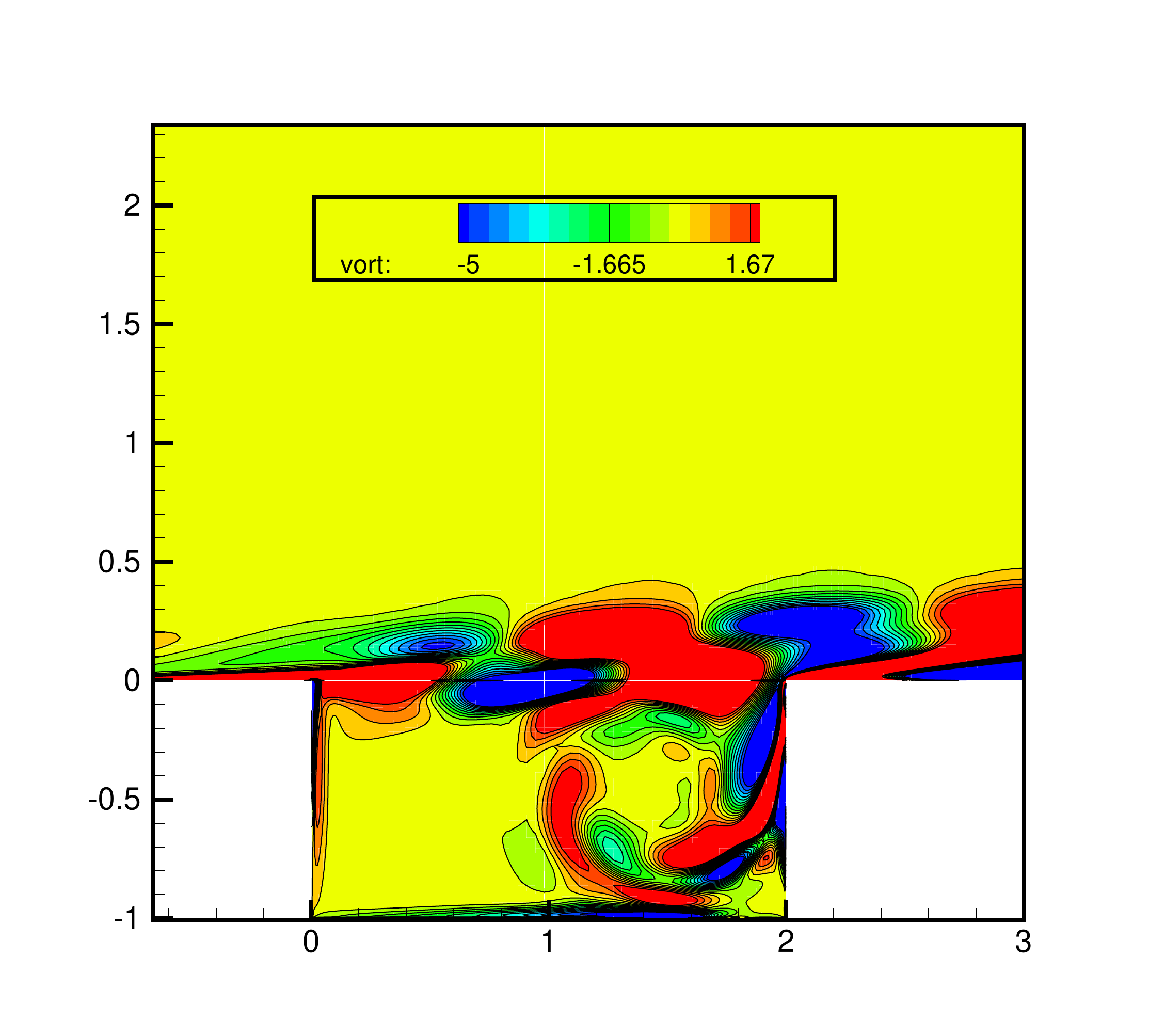}}
\subfigure[mode 2]
{\includegraphics[width=0.45\textwidth]{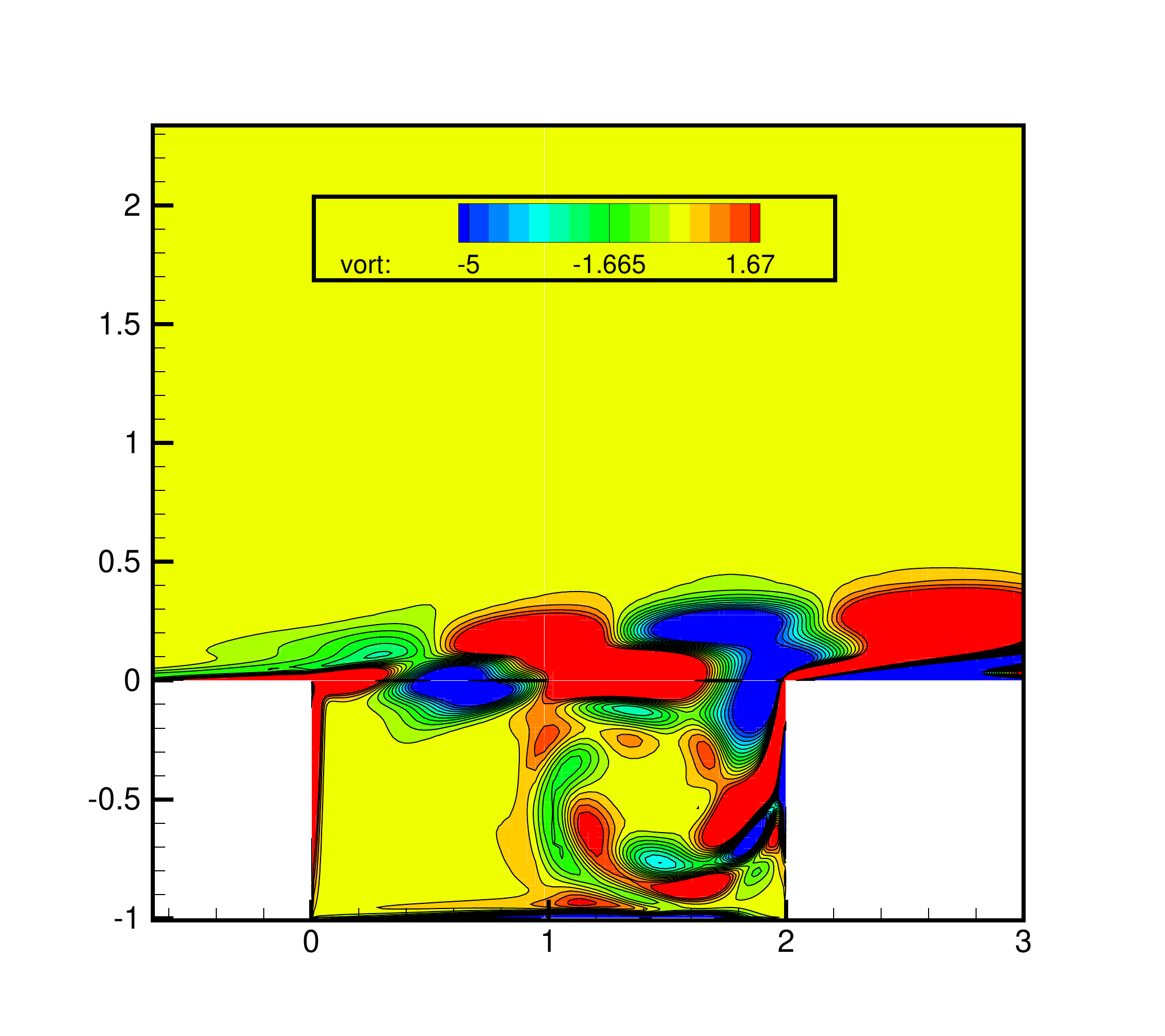}}
\\
\subfigure[mode 3]
{\includegraphics[width=0.45\textwidth]{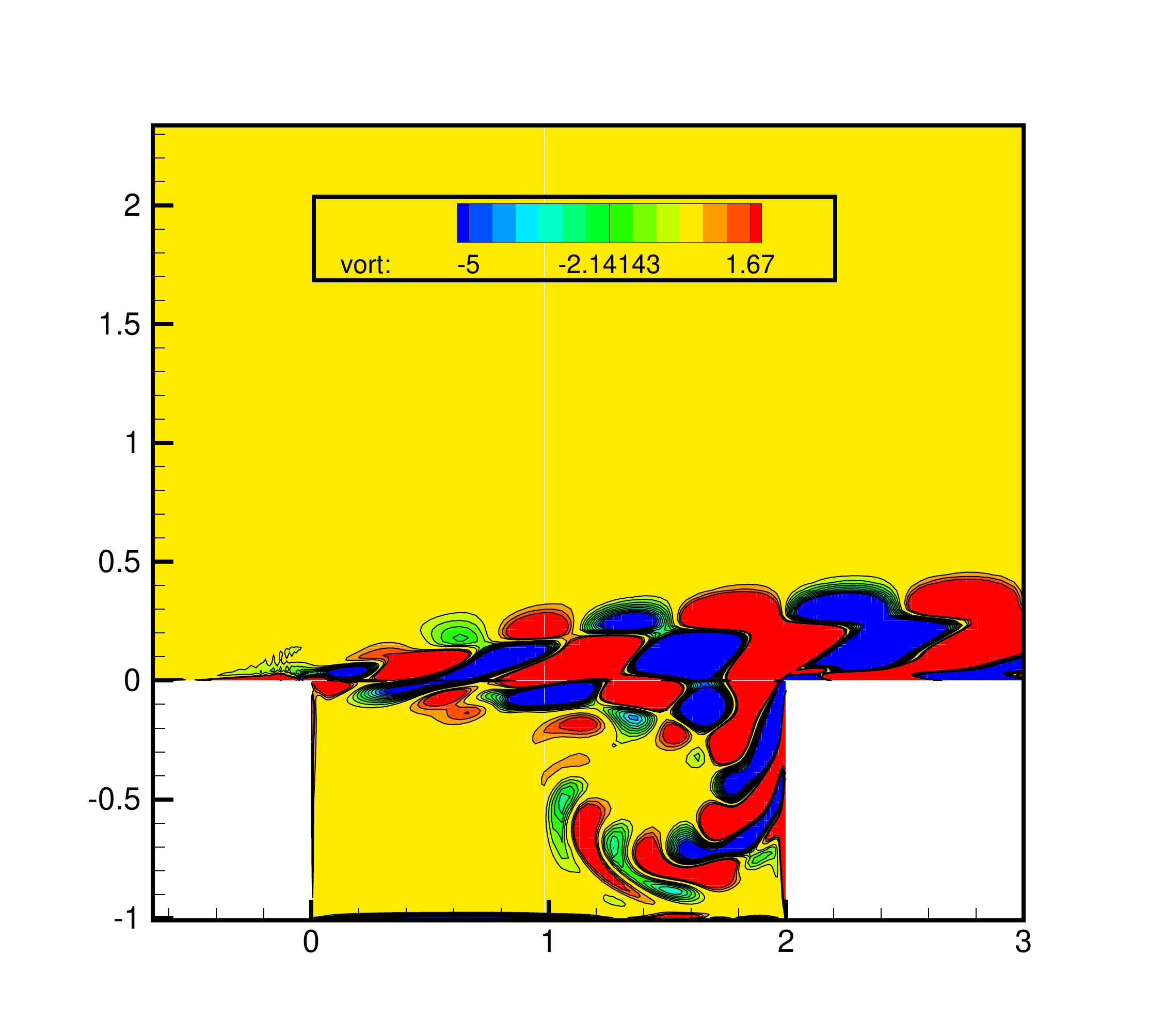}}
\subfigure[mode 4]
{\includegraphics[width=0.45\textwidth]{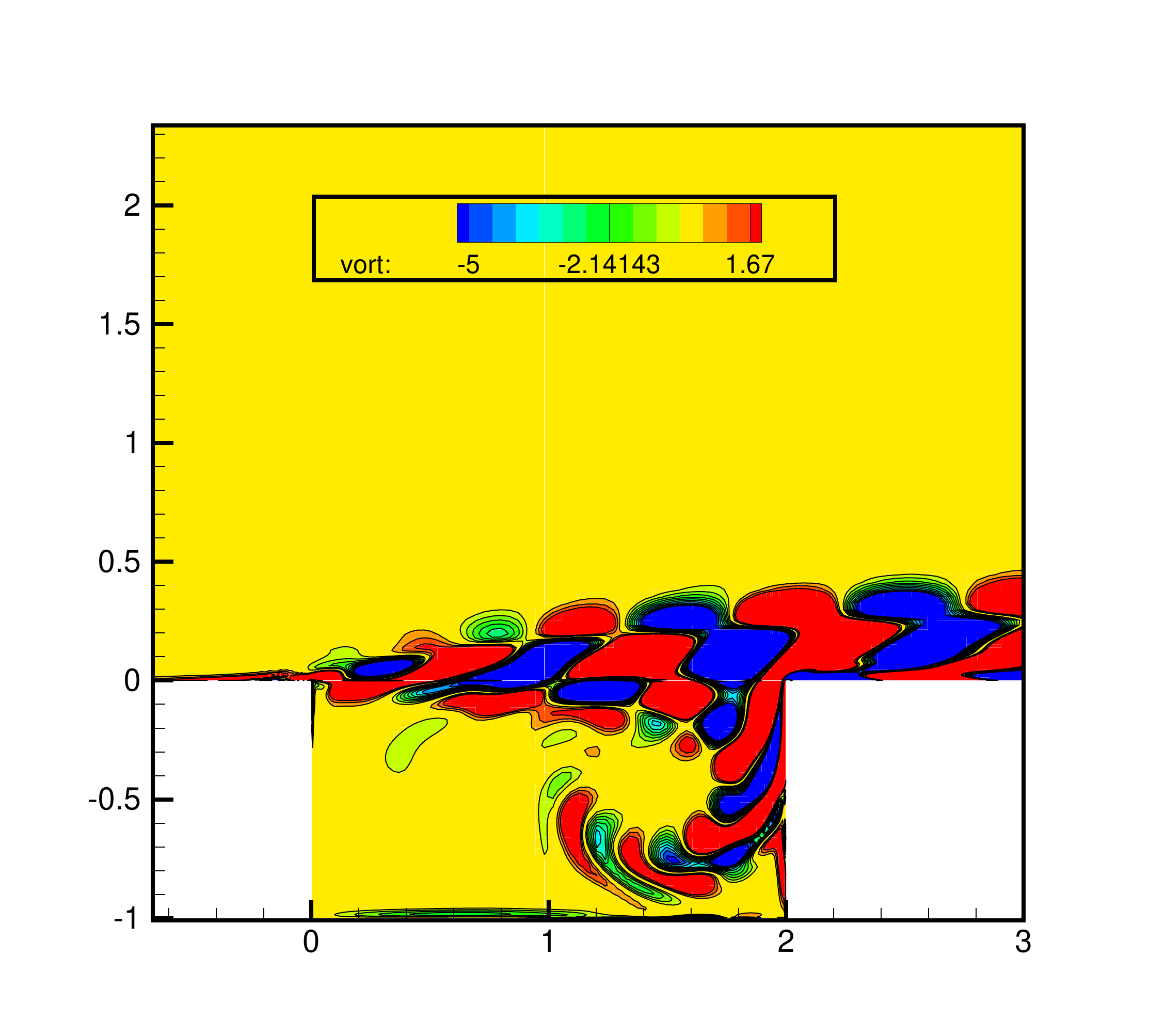}}
\caption{Vorticity contours of the first $4$ POD modes. $15$ contours in the range $[-5,1.67]$ are plotted.}
\label{figcicp:vortmodes}
\end{figure}

\begin{figure}
\centering
\subfigure
{\includegraphics[width=0.4\textwidth]{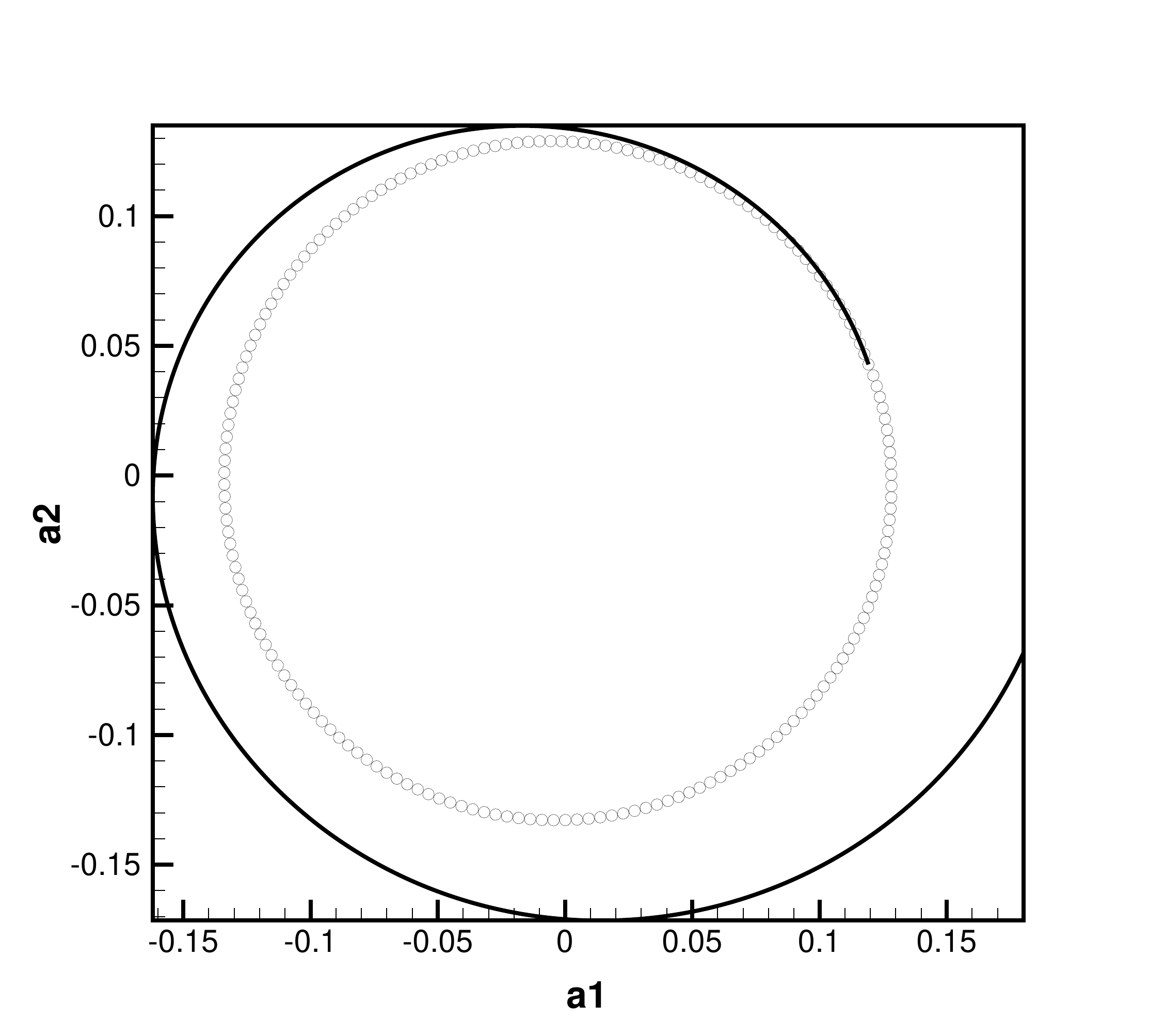}}
\subfigure
{\includegraphics[width=0.4\textwidth]{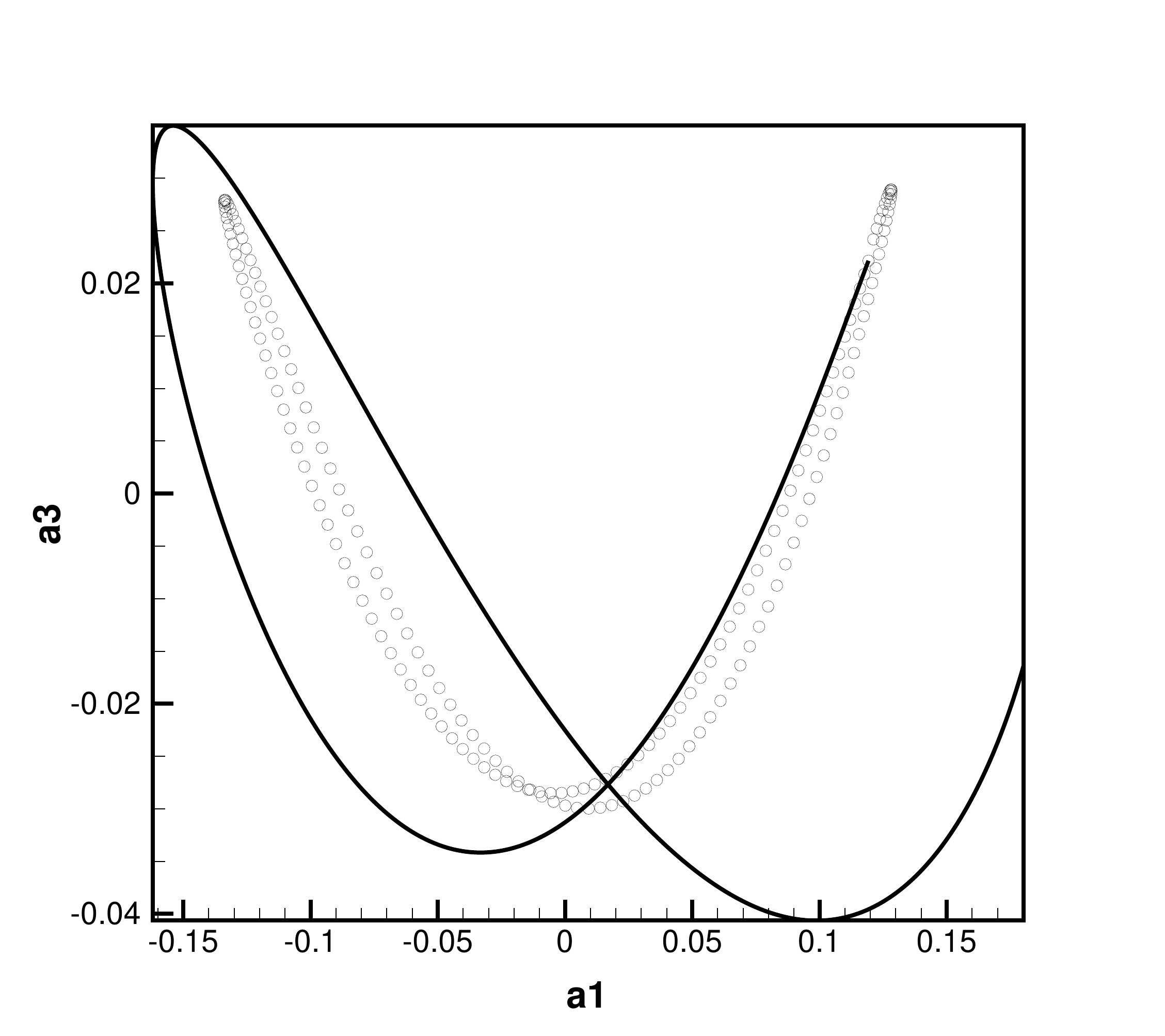}}
\caption{Phase Portraits of mode $a_1$ with mode $a_2$ and mode $a_1$ with mode $a_3$ for GP-ROM predictions without Calibration. GP-ROM: (solid line), POD: (black $o$)}
\label{fig:phasePortraitNoCalib}
\end{figure}

Snapshots are taken once the flow has stabilised and  $56$ snapshots are uniformly sampled which corresponds to  $1$ period of the flow oscillation ($2.75$ in non dimensional time) corresponding to the first Rossiter mode~\citep{delprat2006rossiter}. Figure \ref{figcicp:ric1} demonstrates a degenerate eigen-spectrum showing eigenvalues $\sigma_j$ which occur in pairs.  Also the first 4 eigenmodes capture around $98.5\%$ of the total fluctuation energy as shown by the Relative Information Content (RIC)  defined as $RIC(i)=\sum_{j=1}^i\sigma_j/\sum_{j=1}^M \sigma_j$, where $M$ is the total number of eigenvalues. The temporal coefficients for the dominant $6$ POD modes are shown in figure \ref{figcicp:temporalcoeff}, which displays a phase difference of $\pi/2$ between the paired modes. 
The representation of the first 4 spatial POD modes for the vorticity is shown in figure \ref{figcicp:vortmodes} where they occur in pairs. While their values are distinct, we note that their representation is topologically equivalent. These topological features are representative of the low frequency dynamics of the flow vorticity, which is a hydrodynamic phenomenon of interest. The key issue plaguing  GP ROM in Eqn.~\ref{gpEqn} is its intrinsic tendency to converge to the wrong attractor when used to simulate the flow system for long time periods~\citep{amsallem2012stabilization,barone2009stable,huang2016investigation}. This is clearly seen in Figure~\ref{fig:phasePortraitNoCalib} which shows the phase portraits of the GP-ROM predicted POD temporal coefficients $a_1$ with $a_2$, and $a_1$ with $a_4$. We can observe that the phase portraits quickly become unstable, as a result of errors in $a_i$, thereby creating significant barriers in using the GP-ROM for feedback control, since they require \textit{stable} predictions over some extended time horizon.

\begin{figure}
\centering
\subfigure
{\includegraphics[width=0.4\textwidth]{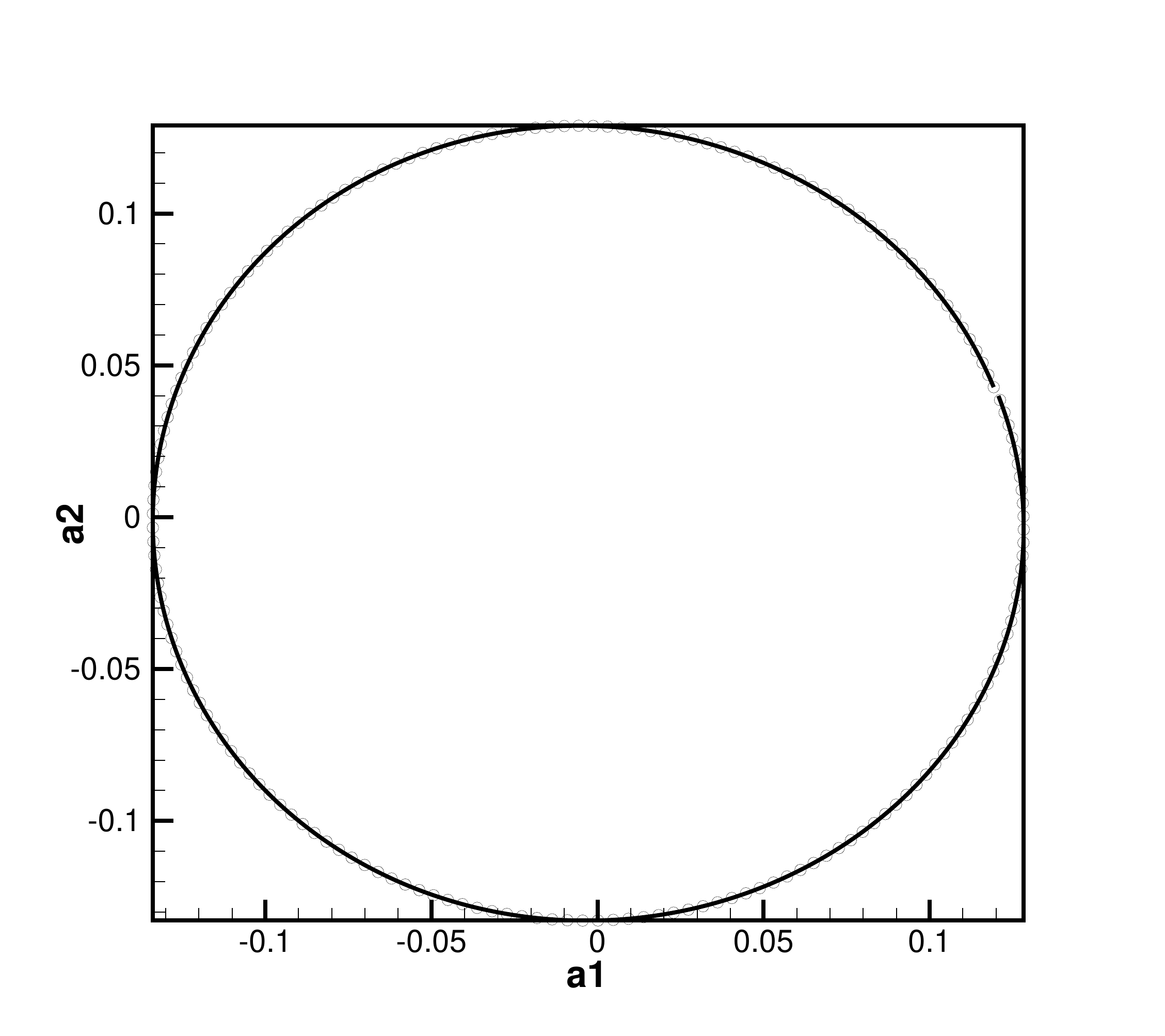}}
\subfigure
{\includegraphics[width=0.4\textwidth]{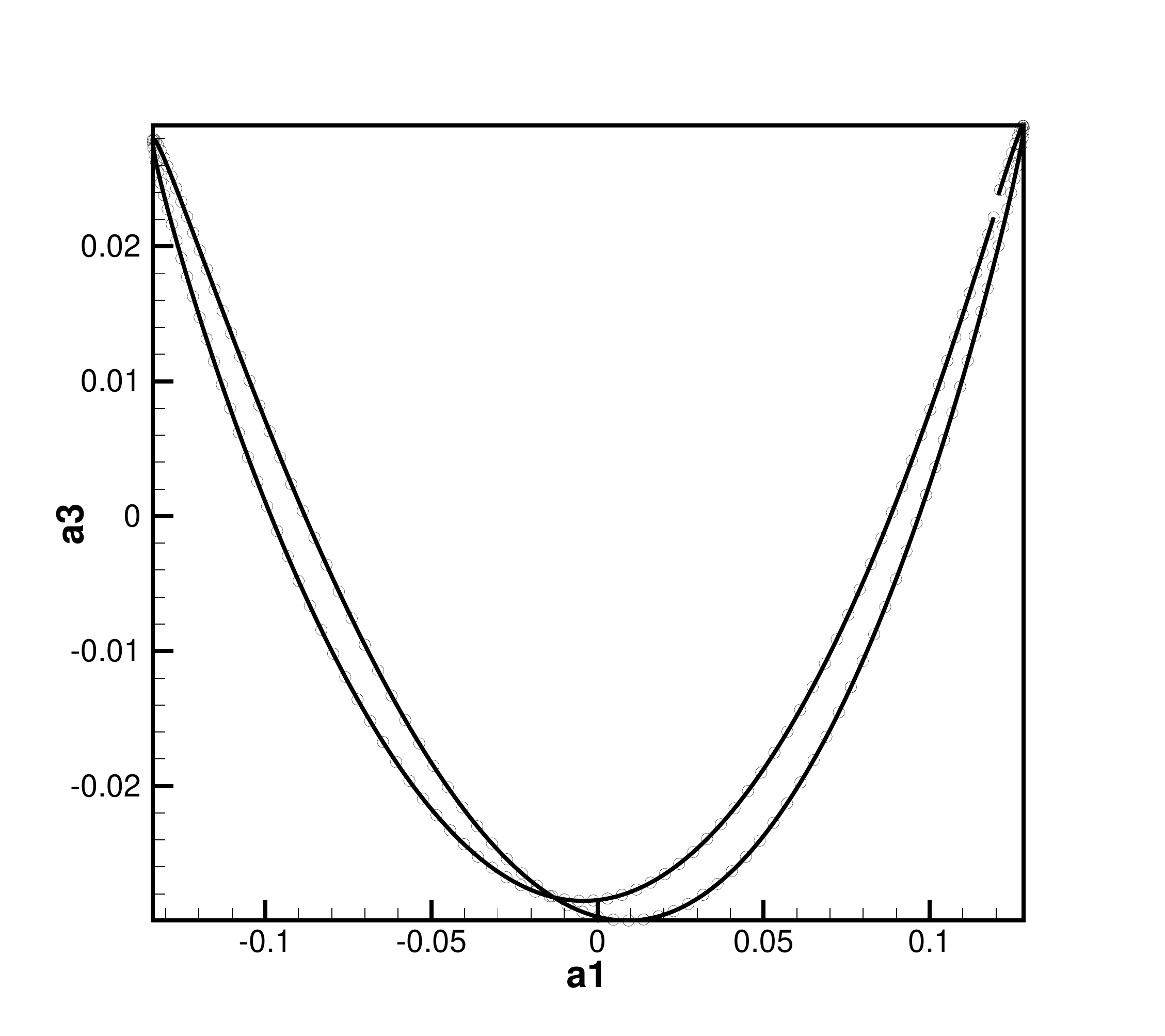}}
\caption{Phase Portraits of mode $a_1$ with mode $a_2$ and mode $a_1$ with mode $a_3$ for GP-ROM predictions with Calibrated Coefficients. GP-ROM: (solid line), POD: (black $o$)}
\label{fig:phasePortraitWithCalib}
\end{figure}

\begin{figure}
\centering
    \includegraphics[width=11.0cm]{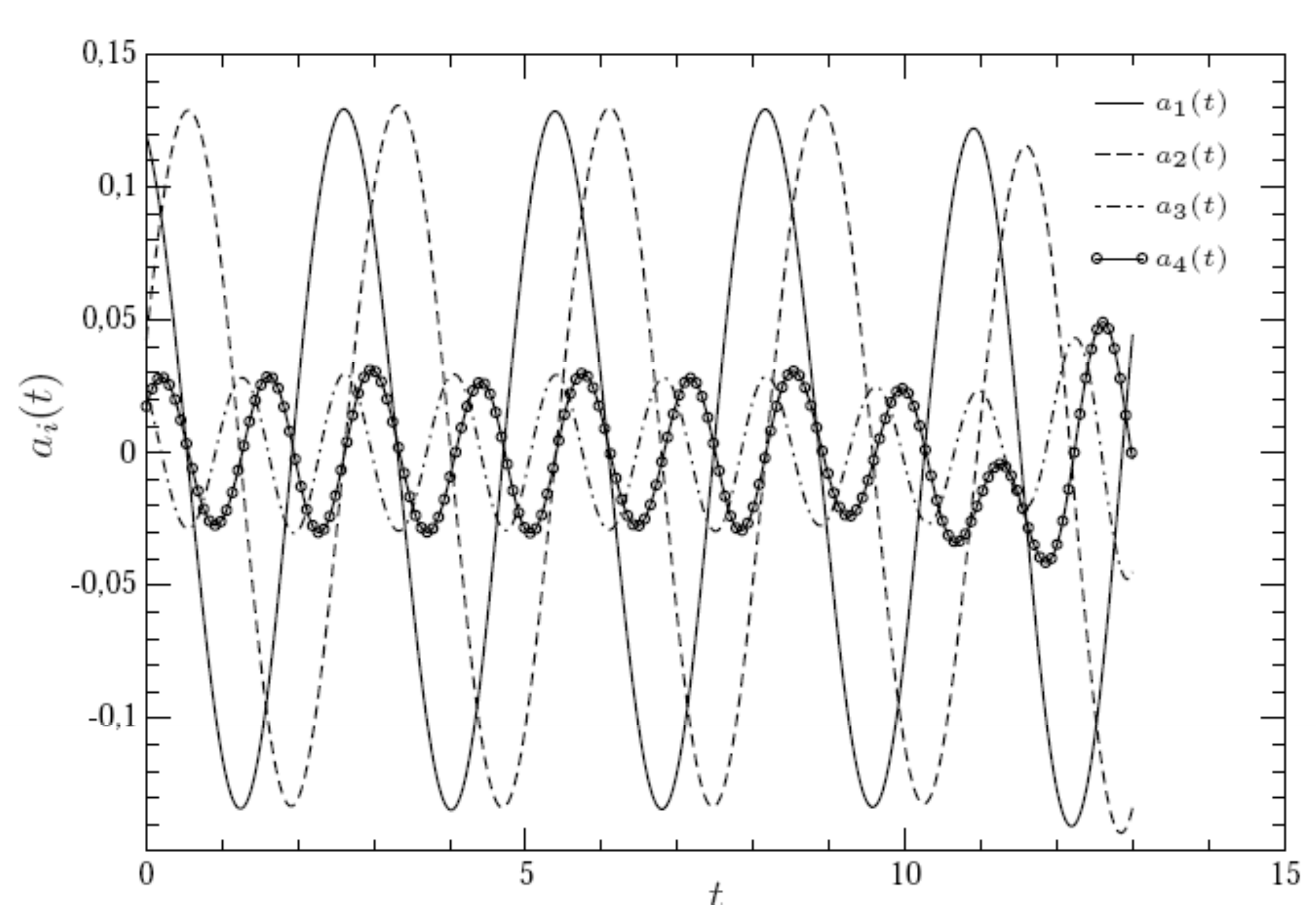}
    \caption{GP-ROM predictions of $a_{i}(t)$ with Calibrated Coefficients}
\label{GPROMpredictions}
\end{figure}

A major reason for the inaccurate behavior of the ROM can be attributed to the truncation of the POD bases where the dissipative scales of the higher POD modes are neglected. An analogous problem occurs in the Large Eddy Simulation (LES) of flows where there is lack of dissipation due to the smaller scales.  Even including all the modes in Galerkin projection may still lead to the wrong attractor due to structural instability as has been demonstrated in \cite{rempfer2000low}.  Other problems may arise due to the contribution of pressure at the boundaries of the domain,  which is usually neglected \citep{noack2005need}. The stability properties of the compressible POD-Galerkin approximation has been studied by \cite{iollo2000stability}. Hence, there is a need to identify the temporal coefficients so as to minimize the error between the actual time coefficients $a_i^{P}(t)$ from the POD and that obtained from the ROM $a_i^{R}(t)$ using a suitable norm for the error. A typical solution in literature is to add an additional ``calibration" step over the GP-ROM. In this step, we have to find the coefficients of the dynamical system such that the error $\bm{e}^{1}$ is minimized under the constraints that the coefficients $C_i, \bm{L_i}, \bm{Q_i}$ ($i=1,\cdots,n$) satisfy Eqn~\ref{rom}, where  $\bm{e}^{1}(\bm{f},t) = \bm{a}^{P}(t) - \bm{a}^{R}(t)$. This leads to a non-linear constrained optimization problem which solved with weighted Tikhonov regularization, the details of which are presented in Appendix and in~\cite{nagarajan2013development}. The phase portraits for the same coefficients in Fig.~\ref{fig:phasePortraitNoCalib} \textit{after calibration} are shown in Fig.~\ref{fig:phasePortraitWithCalib}, demonstrating stable predictions for the individual coefficients. This provides confidence in the calibration, and we now make predictions for long time horizons, as shown in Fig.~\ref{GPROMpredictions}. Since we are dealing with statistically stationary flow it is natural to expect  the temporal dynamics be valid for a time longer than the period of snapshot acquisition i.e. one cycle. As shown by \cite{sirisup2004spectral}, even if the system is initialized with correct state, the solution may drift away for a long period of integration despite calibration. Figure \ref{GPROMpredictions} shows that the calibrated model predicts $a_{i}(t)$ for a non-dimensional time $t \,=\, 11$, which corresponds to around $4$ cycles of the flow period. While this is a marked improvement over uncalibrated coefficients, we notice that the predictions still diverge rapidly after $4$ cycles. The reason is that the neglected modes which contribute to the regularization of the system are not modeled during calibration. The validity of ROM for a longer time of integration is still an open question, since the coefficients need to be able to model the neglected modes. While performing control, since the time period where the control law is applicable is much larger than the period of validity of the model, we calibrate the coefficients for more than one cycle of the flow.

\subsection{Neural Galerkin Projection ROMs of Cavity Flow}
\label{sec:results:neuralGP} 
We now present results for the NeuralGP approach, as applied to the cavity flow problem desribed above. The NeuralGP ODE is constructed consistent with Eqn.~\ref{neuralGP}, where the $C$ and $L$ terms are represented by fully connected NNs taking $f(q)^{C}$ and $f(q)^{C}$ respectively, as input. Since the NN learns these terms based on the relationship between $f(q)$ with $\bm{a}^{P}(t)$, we have a choice of describing $f(q)$. In this work, we choose $f(q)^{C}$ and $f(q)^{L}$ to be non-zero random matrices. Using random matrices as input to the NeuralGP is analogous to the generative learning problem, such as the ubiquitous GANs~\citep{goodfellow2014generative} where a latent random matrix serves as a ``noise" vector that the NN samples to learn a probability distribution. As such, it is important we save these matrices, since it will serve as the input seed for the trained NeuralGP network when making predictions. We note that the input can be any physically motivated quantity as well, if it is available.

To prototype the validity of the NeuralGP method, we attempt to learn from a training data $\bm{a}^{P}(t)$ corresponding to one cycle $ t \approx 2.75 s$ and predict the same cycle. On the surface, this seems like a contrived case since the learning process is considerably easier due to likely over-fitting of the training data. However, the goal of the prototyping phase is to create a ``sanity" test to check if the NN architecture and the learning problem are setup to learn parameters of interest to making predictions. It is entirely possible to construct NNs that are incapable of accurate predictions due to the learning problem being set up in an ambiguous and inconsistent manner, irrespective of the computing horsepower employed. A generally accepted heuristic sanity check for NNs is to attempt deliberate over-fitting of the data, as it is the simplest learning problem for the NN; compared to the extremely challenging problem of generalization and prediction on out-of-sample test data. If we find that the predictions are accurate, it gives us reason to be optimistic about out-of-sample prediction by avoiding over-fitting. Conversely, failure in this contrived problem may indicate fundamental limitations in the NN architecture or learning problem formulation, which need to be addressed before proceeding to harder problems. In this work, where we aim to predict time series, the training data is one cycle of the flow in the cavity and the out-of-sample test data is the subsequent cycles in the future. Since the time series is from a stationary dynamical system, all cycles are \textit{qualitatively} similar, but \textit{quantitatively} different, requiring that the NN learn its statistical properties to generalize.

\begin{figure}[ht]
\centering
\subfigure
{\includegraphics[width=0.45\textwidth]{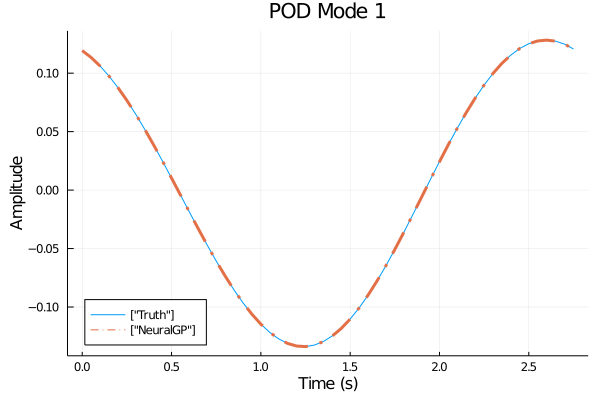}}
\subfigure
{\includegraphics[width=0.45\textwidth]{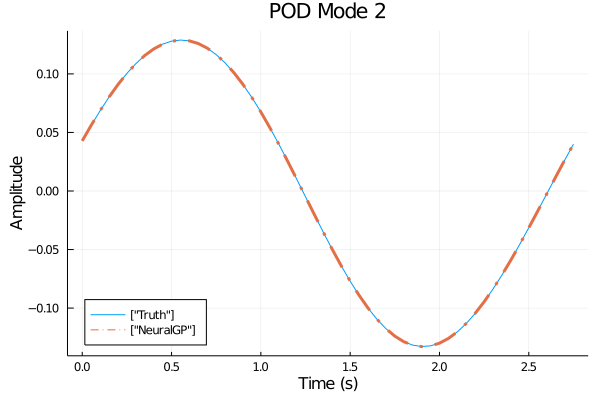}}
\subfigure
{\includegraphics[width=0.45\textwidth]{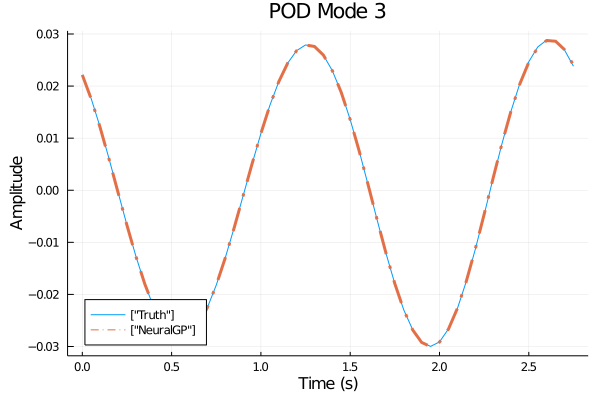}}
\subfigure
{\includegraphics[width=0.45\textwidth]{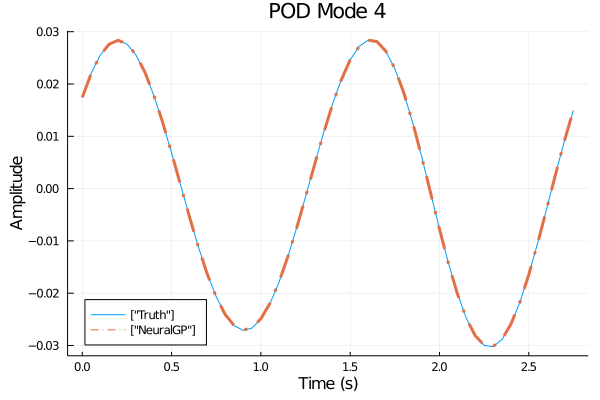}}
\subfigure
{\includegraphics[width=0.45\textwidth]{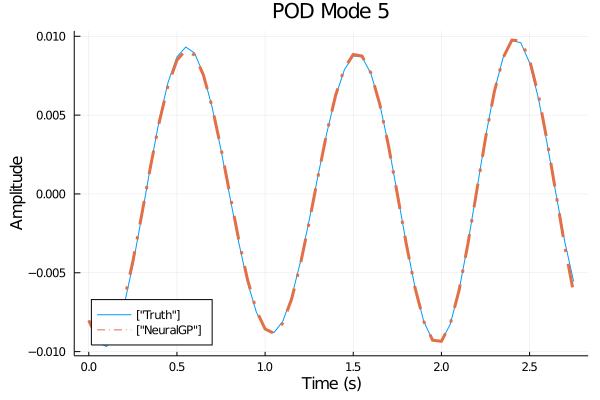}}
{\includegraphics[width=0.45\textwidth]{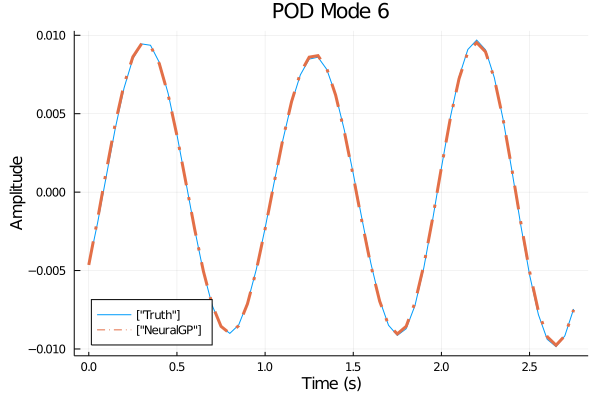}}
\caption{NeuralGP-ROM predictions of $\alpha_{i}(t)$ for $1$ cycle}
\label{fig:1cyclePrediction}
\end{figure}

\begin{figure}[ht]
\centering
\subfigure
{\includegraphics[width=0.45\textwidth]{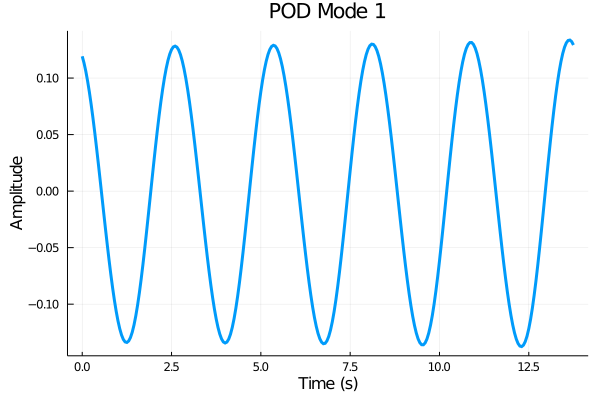}}
\subfigure
{\includegraphics[width=0.45\textwidth]{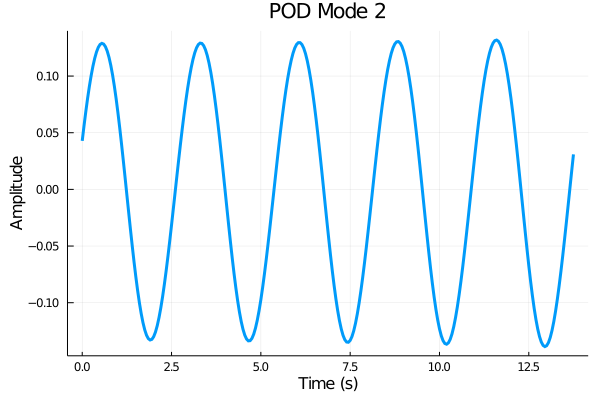}}
\subfigure
{\includegraphics[width=0.45\textwidth]{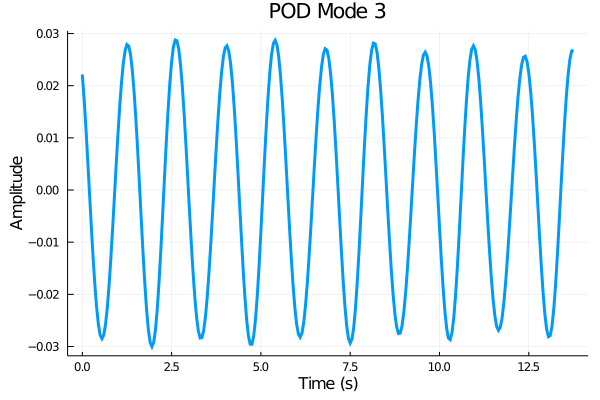}}
\subfigure
{\includegraphics[width=0.45\textwidth]{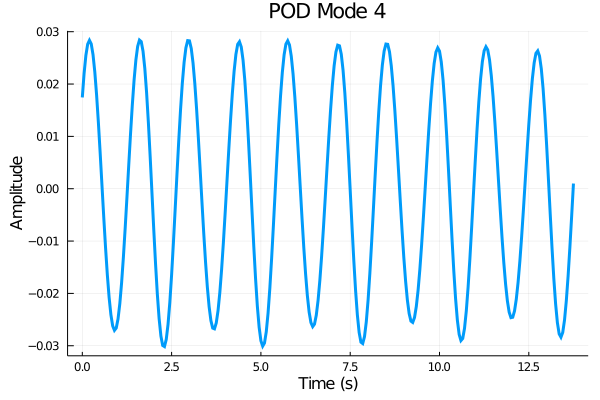}}
\subfigure
{\includegraphics[width=0.45\textwidth]{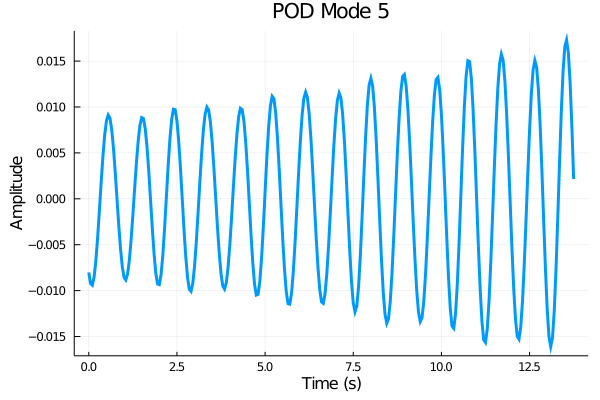}}
{\includegraphics[width=0.45\textwidth]{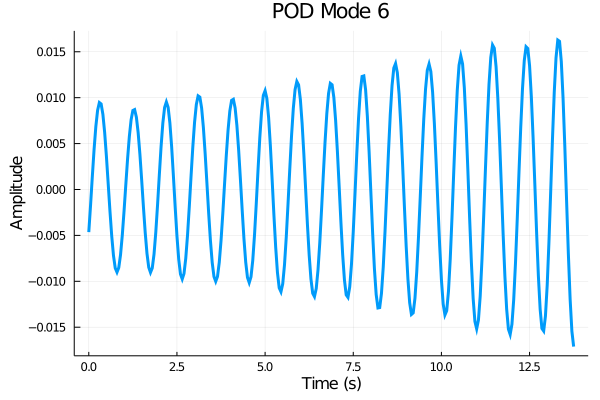}}
\caption{NeuralGP-ROM predictions of $\alpha_{i}(t)$ for $5$ cycles}
\label{fig:5cyclePrediction}
\end{figure}

Figure~\ref{fig:1cyclePrediction} shows the results of prototyping sanity test, with the predicted and the training $\bm{a}(t)$ plotted for the $6$ dominant POD modes. We see that the predictions are extremely good, with an almost perfect match with the training data. This indicates that the NeuralGP indeed are capable of learning the data with great accuracy, and thus provides confidence that our architecture and learning problem are appropriate. With this validation, we proceed to the next step which is of real consequence to engineering problems, namely the prediction of cycles beyond the training horizon. To evaluate this predictive capability, we again train NeuralGP with one cycle as shown above, and predict $5$ cycles ahead. This is representative of flow control applications where the system ROM predicts several cycles in advance, for the control algorithms that construct actuation responses to keep the flow in the desired state. The prediction results for the $6$ dominant $\bm{a}(t)$ are shown in Fig.~\ref{fig:5cyclePrediction} where $5$ cycles $\approx 12.75 s$. Here we notice that the predictions of modes 1-4 are remarkably stable for almost the entire horizon. In cases of modes 5 and 6, the instability propagates after $\approx 5 s$, which corresponds to almost two cycles. It is important to note that unlike the standard GP ROM, we need not perform any explicit manual calibration of the coefficients. Instead, the NeuralGP automatically learns the best coefficients that are consistent with both the data and the GP equation structure. To provide further context, GP-ROMs of dynamical systems place heavy emphasis on the accuracy of the most significant $\bm{a}(t)$ (ranked by eigenvalues) as they have a greater impact on the overall ROM and control system than the other modes. In fact, a successful GP-ROM based feedback control system for the cavity problem in this work was previously demonstrated in~\cite{nagarajan2013development} with just POD modes 1-4. From this perspective, we observe that NeuralGP-ROM can easily exceed performance and accuracy metrics already established for control applications, with minimal intervention. 

\begin{figure}[ht]
\centering
\subfigure
{\includegraphics[width=0.45\textwidth]{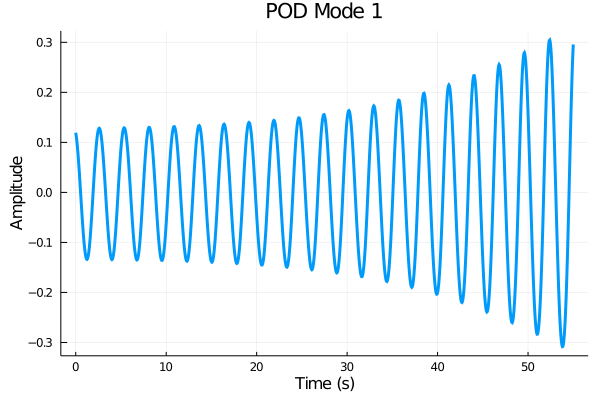}}
\subfigure
{\includegraphics[width=0.45\textwidth]{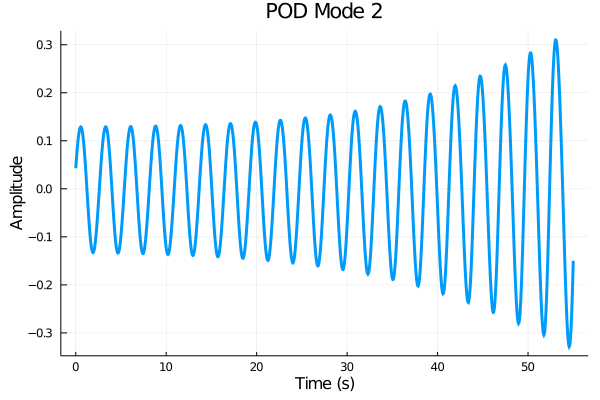}}
\subfigure
{\includegraphics[width=0.45\textwidth]{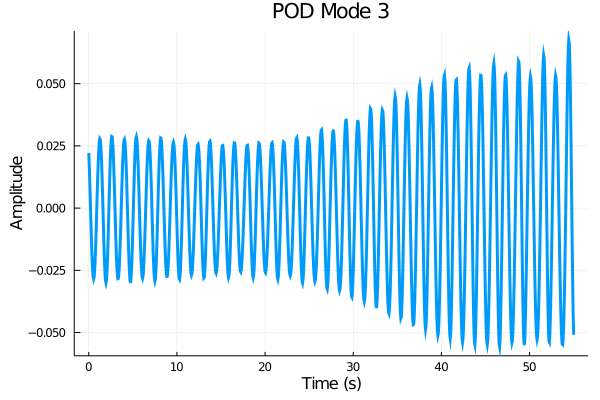}}
\subfigure
{\includegraphics[width=0.45\textwidth]{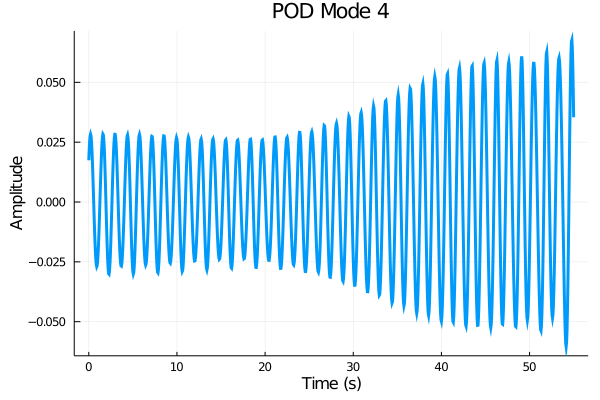}}
\subfigure
{\includegraphics[width=0.45\textwidth]{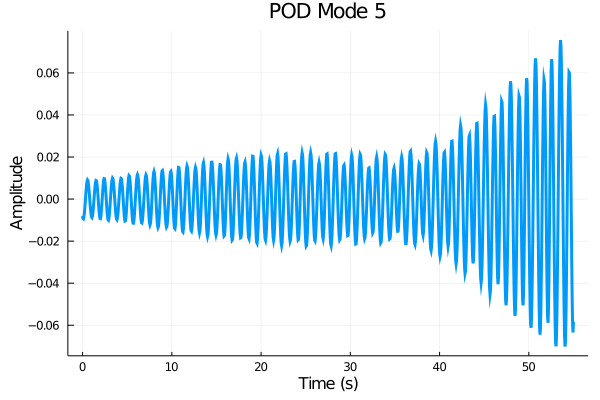}}
{\includegraphics[width=0.45\textwidth]{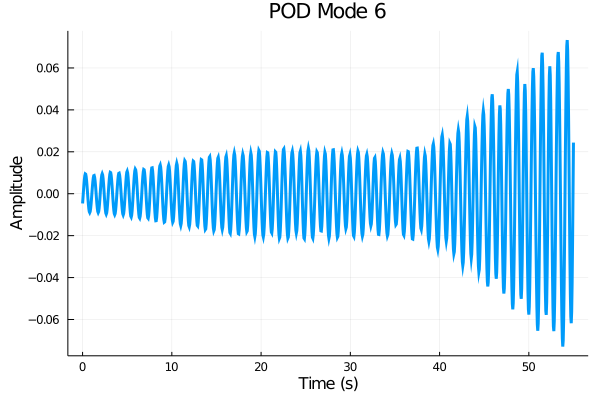}}
\caption{NeuralGP-ROM predictions of $\alpha_{i}(t)$ for $20$ cycles, with divergence observed at $\approx$ 10 cycles}
\label{fig:20cyclePrediction}
\end{figure}

At this juncture, we remark that turbulence ROMs by virtue of construction - whether data-driven, physics-based or a hybrid combination (like the NeuralGP) - typically neglect some physics of the dynamical system they seek to represent. In case of complex, chaotic systems like turbulence, the challenge is to choose which physical features to retain in the ROM, and the features to neglect/truncate. While this truncation is necessary to construct efficient, parsimonious ROMs, it causes instabilities in the long time evolution of the dynamical system. As a result, any truncated ROM is eventually expected to become inaccurate and degenerate into an unstable solution at some point in its evolution. 
As a result, a key metric of evaluation is \textit{when} the ROM destabilizes as this acts as a measure of its quality. We aim to study this for NeuralGP-ROM, noting from Fig.~\ref{GPROMpredictions} that the existing GP-ROM with calibration destabilizes at $t \approx 11 s$. To this end, the trained NeuralGP is instead used to predict $20$ cycles i.e. $55 s$ into the future for all $6$ dominant $\bm{a}(t)$, and the results are presented in Fig.~\ref{fig:20cyclePrediction}. Interestingly, we observe remarkably stable predictions for POD modes 1 and 2 until $t \approx 28 s$, which corresponds to $\approx 10$ cycles before rapidly diverging. This trend is also seen in POD modes 3 and 4, where stability is reached for as long as $9-10$ cycles before diverging. As expected from the previous results, the prediction is weaker for modes 5 and 6, where stability is observed for $\approx 3$ cycles, after which it destabilizes - first gradually and then exponentially further into the horizon. While this may seem unimpressive at first glance, we reiterate that stable ROM based controls have been demonstrated for this problem with only modes $1-4$ in~\cite{nagarajan2013development} as they capture a significant percentage of the flow structures. 

Nevertheless, the inclusion of the higher modes $5$ and $6$ in this work is to illustrate a key challenge in predicting higher modes with similar levels of accuracy. A major reason is the large variance in amplitudes of these $6$ modes relative to each other. From Fig.~\ref{fig:1cyclePrediction}, we see that the amplitude of mode $6$ is almost two orders of magnitude smaller than mode $1$, while mode $3$ is itself an order of magnitude apart from both of them. This aspect is crucially relevant due to the construction of the loss function in NNs, since it is an aggregate quantity summed or averaged over all the output quantities in any given batch. For example, in RMS loss function for $N$ modes (where $N=6$), $ \sum_{i=1}^{i=N} a_{i}^R(t) - a_{i}^{train}(t)$, the discrepancy between the predicted and training data is summed over to a scalar quantity. This is necessary since the standard form of reverse mode automatic differentiation; for \textit{backpropagation} in NNs needs a scalar value to compute derivatives. As a result, the scalar loss has more contribution from the modes with relatively larger magnitude (modes $1-4$) than modes $5-6$ and the NNs learn coefficients that prioritize the dominant modes. Future work will explore methods to improve accuracy in higher modes by modifying the structure of the Eqn.~\ref{neuralGP}, for applications that require superior resolution of higher modes. However, since most engineering applications emphasize dominant modes, the results demonstrate superiority of the NeuralGP approach, since its prediction horizon is stable for \textit{three times longer} than even the calibrated GP-ROM without any additional training dataset, or manual parameter tuning.

\section{Conclusion}
\label{sec:conclusion} 
It has been widely reported in literature that predictions from Galerkin-Projection based reduced order models (ROMs) of turbulence suffer from lack of long-term stability and desired accuracy, despite the approach being theoretically derived as a data-driven approximation of the Navier-Stokes equations. This can be a serious issue, since these ROMs are used for a variety of applications ranging from feedback control to surrogate modeling for rapid decision making. Despite the success of machine learning (ML) based ROMs in the recent years, its black-box nature has raised concerns about interpretability and confidence in its predictions. In this work, we have proposed a hybrid ROM blending both ML and Galerkin projection  - coined the \textit{Neural Galerkin Projection} (\textit{NeuralGP}), by expressing the Galerkin projection onto the Navier Stokes equations as a Differentiable Programming problem. The NeuralGP learns the coefficients of the Galerkin projection ODE directly from data by embedding NNs inside the ODE. The  NNs can be standard dense layers with user defined choice of parameters, learning rates and optimizers, while still adhering to the physics of the Galerkin ODE. This is in contrast to the existing Galerkin ROMs, which require manual calibration of the coefficients from the data. The results show that in contrast to a fully ML based ``black-box" ROM, the NeuralGP-ROM is extremely interpretable with its strong physics foundations, while being more accurate than traditional Galerkin ROMs. Specifically, predictions demonstrate that NeuralGP-ROMs indeed generalize better and can predict as long as $10$ cycles into the future, despite training on only a single cycle. Furthermore, the computational cost of training is extremely low since retention of the Galerkin ODE structure avoids the need of expensive, complex deep NN architectures like LSTMs and CNNs as black boxes for the entire ODE which requires several GPUs to train~\citep{mohan2018deep,mohan2019compressed,mohan2020spatio}. As a result, training took just over an hour on a desktop grade CPU. An attractive benefit of hybrid ML ROM; despite its theoretical sophistication, is that the underlying equation is still an ODE that can be solved with mature, standard ODE solvers. This allows us to leverage and improve existing ODE-based feedback control infrastructure by seamlessly replacing GP-ROM with NeuralGP-ROM, with almost no changes to the control system elsewhere. The NeuralGP is extremely flexible as it is based on the differentiable programming paradigm, which allows us to add arbitrary terms to the ODE based on our partial knowledge of the physics, while learning the unknown physics with neural networks~\citep{gelbrecht2021neural}. In essence, the NeuralGP is an attempt to bring together the best of both worlds from Galerkin projections and deep learning, and the approach can be further extended to include additional physics constraints and even effects of truncated modes. 

\section{Acknowledgements}
This work has been co-authored by employees of Triad National Security, LLC which operates Los Alamos National Laboratory (LANL) under Contract No. 89233218CNA000001 with the U.S. Department of Energy/National Nuclear Security Administration. A.T.M and D.L. were supported by the LDRD (Laboratory Directed Research and Development) program at LANL under project 20190059DR. Computational resources were provided by the Institutional Computing program at LANL. K.N. has been supported by Aeronautical Research and  Development Board, India, Project number 1912, titled ``Development and application of flow control techniques for flow past a rectangular cavity and circular cylinder". We also thank Chris Rackauckas for assistance with the \textit{DiffEqFlux.jl} package.

\section{Appendix}
\label{app}
\subsection*{Proper Orthogonal Decomposition}
\label{app:POD}
This section expands on the POD representation shown in Eqn.~\ref{PODeqn}, where any given flow variable is expanded into an average component 
$\bm{\overline{q}}$ of the $n$ flow snapshots and a fluctuating component. The fluctuating component is represented by its spatial modes $\bm{\phi_i}(\bm{x})$ and the temporal coefficients $a_i(t)$, which are determined by solving an eigenvalue problem involving the time correlation tensor that minimises the average projection error:
\begin{equation}
 E\left(\Vert \bm{q} - P_S \bm{q} \Vert \right)
\end{equation}
$E$ denotes the averaging operator, for instance the ensemble average $(E = \frac{1}{n}\sum_{i=1}^n)$ and $P_S$ denote the projection operator over the space $S \subset H$ and dimension $m$.
Note that the problem of minimizing $E(\Vert \bm{q} -P_{S} \bm{q} \Vert)$ is equivalent to maximizing $E(\Vert P_{S} \ \bm{q} \Vert^2)$, the "energy" of the orthogonal projection since by Pythogoraus theorem we have $\Vert \bm{q} \Vert^2 = \Vert \bm{q} - P_S \ \bm{q} \Vert^2 + \Vert P \ \bm{q} \Vert^2$.    Solving the optimisation problem leads to an eigenvalue problem for the spatial modes $\bm{\phi}$ given by
\begin{equation} \label{infinite_eigen}
R \bm{\phi} = \lambda \bm{\phi}
\end{equation}
where $R: H \rightarrow H$ is the linear operator  defined as 
\begin{equation}
R  = E(\bm{q}_k \otimes \bm{q}_k^{*})
\end{equation}
Here $\bm{q}^{*}\in H^*$ denotes the dual of $\bm{q}$, given by $\bm{q}^*(.) = \left(.,\bm{q}\right)$ and $\otimes$ is the usual tensor product \footnote{We have $(\bm{u} \otimes \bm{v}^*)(\bm{\psi}) = \bm{u} \left (\bm{\psi} , \bm{v} \right),     \forall \bm{u}, \bm{v}, \bm{\psi} \in H$.}.
One can easily verify that R is self adjoint \textit{i.e.} $\left(R \ x, y\right) = \left(x, R \ y\right)$ and hence by the spectral theorem the eigenfunctions $\bm{\phi}$ can be chosen to be orthonormal,  The eigenvalues $\lambda$ are then determined by taking the inner product of equation (\ref{infinite_eigen}) with $\bm{\phi}$
to obtain
\begin{equation}
\lambda = E\left(\vert \left( \bm{q}_k , \bm{\phi} \right) \vert^2 \right)
\end{equation}
The eigenvalues $\lambda$ in the above equation represents the average energy in the projection of the ensemble onto $\bm{\phi}$
where energy is defined in sense of the induced norm.   We also conclude that the $R$ is positive semi-definite, and the eigenfunctions $\bm{\phi_j}$ which maximise $E(\Vert P_{S} \bm{q}_k \Vert^2)$ are the eigenfunctions corresponding to the largest eigenvalues.   
Also  the eigenfunctions reproduce almost every member of the ensemble (except on a set of measure zero).    One can also verify that the range of the operator $R$ is contained in the span of the ensemble $\bm{q_k}$ and  that any eigenfunction $\bm{\phi}$ can be written as linear combination of snapshots: 
\begin{equation}
\bm{\phi} = \sum_{k=1}^n c_{k}\bm{q}_{k}
\end{equation}
where the coefficients $c_k \in \mathbb R$.  If the average operator $E$ over the snapshots $\bm{q}_k$ is given as
\begin{equation}
E\left(f(\bm{q})\right) = \sum_{k=1}^n \alpha_{k}f(\bm{q}_k)
\end{equation}
where $\alpha_k > 0$ satisfies $\sum_{k=1}^n  \alpha_k = 1$ (usually giving equally weights to snapshots \textit{i.e.} $\alpha_k = 1/m$).  The eigenvalue problem (\ref{infinite_eigen}) may be written in terms of the coefficients $c_k$ as
\begin{equation}
R \ c = \lambda c
\end{equation}
where $c = (c_1, \cdots ,c_m)$ and $R$ is an $m \times m$ correlation matrix with $R_{ij} = \alpha_i\left(\bm{q}_j, \bm{q}_i\right)$. This is the usual method of snapshots as described \cite{sirovich1987turbulence}. Here we note that the direct problem involves  solving an eigenvalue problem possibly on an infinite dimensional space $H$ which may be very large,  on the other hand the method of snapshots involve  solving  only an $m$ dimensional eigenvalue problem  and the method proves computationally efficient if $m$ is small compared with the dimension of $H$.  

\subsection*{GP-ROM Algorithm and Calibration}
\label{app:calibration} 
In this section we describe the method of calibration used in constructing the Galerkin Projection ROM, which employs Tikhonov Regularization. In addition the definition of error $\bm{e}^{1}(\bm{f},t) = \bm{a}^{P}(t) - \bm{a}^{R}(t)$, we can define the error in two more forms a) A state calibration error $\bm{e}^{2}$ without the dynamical constraint given by 
$$\mathbf{e}^2(\mathbf{y},t) = \mathbf{a}^P(t) -\mathbf{a}^P(0) - \int_{0}^T \mathbf{f}(\mathbf{y}, \mathbf{a}^P(\tau))d\tau$$ 
and, b) The flow calibration error $\bm{e}^{3}(\bm{f},t) = \bm{\dot{a}}^{P}(t) - \bm{f}(\bm{a}^{P}(t)) $ where $\bm{f}$ is characterized by the coefficients of Eqn.~\ref{rom}. As demonstrated in \cite{couplet2005calibrated}, minimization of $\mathcal{I}^{3}(\bm{f})=\langle\Vert \bm{e}^{3}(\bm{f},t) \Vert^2\rangle_{T}$ leads to the solution of a  linear system for the coefficients $\mathbf{y}$ defining $\bm{f}$. Unfortunately, this linear system is not well conditioned and leads to numerical divergence when the calibrated coefficients are used to integrate in time \eqref{rom}. For that purpose, \cite{couplet2005calibrated} have introduced a new cost functional defined as a weighted sum of a term measuring the normalized error between the behavior of the model \eqref{rom} with $\bm{f}$ and with the coefficients determined directly by Galerkin projection $\bm{f}^{GP}$ and another term linked to the distance between $\bm{f}$ and $\bm{f}^{GP}$. The value of the weighting factor which represents the cost of calibration was chosen rather arbitrarily and hence was user dependent.  In \cite{cordier2010calibration}, a Tikhonov regularisation method was suggested to improve the conditioning of the linear system. The idea of this method is to seek a regularized solution $\mathbf{y}_\rho$ as the minimizer of the following weighted functional
$$
\Phi_\rho(\mathbf{y})=\|A\mathbf{y}-\mathbf{b}\|_2^2 + \rho \|\mathcal{L}\left(\mathbf{y}-\mathbf{y_0}\right)\|_2^2,
$$
where the first term corresponds to the residual norm, and the second to a side constraint imposed on the solution. $\mathcal{L}$ represents the discrete approximation matrix of a differential operator of order $d$ and $\mathbf{y_0}$ a reference solution. The value $\rho$ is chosen so as to compromise between the minimisation of the norm of the residual for the linear system and the semi-norm of the solution. Here, the regularisation parameter $\rho$ is determined by the classical L-curve method, as described in \cite{cordier2010calibration}. We have the  error defined for the minimisation of the functional $\mathcal{I}^{i}$ as
\begin{eqnarray*}
 \mathcal{I}^{i}(\mathbf{y}) = \left \langle \Vert \mathbf{e}^i(\mathbf{y},t)\Vert_\Lambda^2 \right \rangle_{T}
\end{eqnarray*}
 Usually the matrix $\Lambda$ is chosen as Identity, which means that we give equal weights to all the modes in the definition of calibration. However  this matrix can be chosen in suitable way so as to include the effect of mode selection in the definition. Two ways of defining the weights can be proposed:
\begin{enumerate}
\item \label{eig} We consider that the main interest is in modelling the effect of the energetic structures and hence the eigen-spectra themselves serve as a measure of the relative importance of the modes, which is the most natural choice of the weights for the definition of error.
\item \label{sens}  The error can be based on an overall sensitivity of the model with respect to a cost functional.
\end{enumerate}
The weight matrix $\Lambda$ for the definition of error for the first case  can be simply written as a diagonal matrix:
$$\Lambda_{ii} =  \frac{\sigma_i}{\mbox{max} \ \sigma_i} \hspace{0.5cm} \mbox{for} \hspace{0.5cm} i=\{1, \cdots, N \} $$ where $\sigma$ is the singular value obtained as a solution of the eigenvalue problem of the time correlation matrix. For the second case consider the state equations
\begin{equation} \label{eq:state}
 \mathbf{\dot{a}}^R = \mathbf{f}(\mathbf{y}, \mathbf{a}^R)
\end{equation}
Variation of any convex cost functional $\mathcal{J}$ with respect to the state variables $\mathbf{a}^R= \{a_i^R\}_{i=1}^N$ gives the adjoint equation of \eqref{eq:state} as
\begin{equation}\label{eq:adj}
 \boldsymbol{ \dot \xi}^R = \mathbf{g}(\mathbf{y}, \mathbf{a}^R, \boldsymbol{{\xi}}^R)
\end{equation}
 Where $\boldsymbol{\xi}^R(t) = \{\xi_i^R\}_{i=1}^N$ is the adjoint variable.  The overall sensitivity of the cost functional $\mathcal{J}$ with respect to $\mathbf{a}^R$ is obtained as
\begin{equation}
  \mathcal{S} =\frac{d\mathcal{J}}{d\mathbf{a}^R} = \langle\mathbf{a}^R(t) \boldsymbol{\xi}^R(t) \rangle
\end{equation}
where $\langle . \rangle$ is any time averaging operator. We can then define the weight matrix $\Lambda$ with respect to the sensitivity as
$$
\Lambda_{ii} = \frac{\mathcal{S}_i}{\mbox{max} \ \mathcal{S}_i} \hspace{0.5cm} \mbox{for} \hspace{0.5cm} i=\{1, \cdots, N \}
$$
In this study we have taken the cost functional $\mathcal{J}$ based on the energy of the temporal modes as
\begin{equation} \label{eq:energyfunc}
 \mathcal{J} = \frac{1}{2} \int_{0}^T \sum_{i=1}^{N} ({a}_i^R(t))^2  dt
\end{equation}
The above functional is minimised subject to the constraint \eqref{eq:state},  by the method of Lagrange multipliers with the adjoint state equation of \eqref{eq:adj}  given by
\begin{equation}\label{eq:adjoint}
 \boldsymbol{\dot{\xi}}(t) = - a_i(t)-\sum_{i=1}^{N}\left({L}_{ij} +\sum_{i=1}^{N}\left(Q_{jik}+Q_{jki}\right)a_k(t) \right)\xi_j(t)
\end{equation}

\bibliographystyle{unsrtnat}
\bibliography{references}  






\end{document}